\documentclass[
 reprint,
 prd,
 nofootinbib,
 amsmath,amssymb,
]{revtex4-2}
\usepackage[english]{babel}
\usepackage{graphicx}
\usepackage{dcolumn}
\usepackage{bm}

\newcommand{\refeq}[1]{Eq.\,(\ref{#1})}
\usepackage{xcolor}
\usepackage{cancel}

\usepackage{hyperref}

\begin{document}

\preprint{APS/123-QED}

\title{$r$-Largest Order Statistics for Galaxy Cluster Cosmology I: Reconstruction of the high-mass end of the Halo Mass Function }

\author{Théo Gayoux}
\email{Theo.Gayoux@obspm.fr}
\affiliation{Laboratoire d’étude de l’Univers et des phénomènes eXtremes, Observatoire de Paris, Université PSL, Sorbonne Université, CNRS, 92190 Meudon, France}
\affiliation{Universit\'e Paris-Cit\'e, 5 Rue Thomas Mann, 75013, Paris, France}
\author{Pier Stefano Corasaniti}%
 \email{Pier-Stefano.Corasaniti@obspm.fr}
\affiliation{
 LUX, Observatoire de Paris, Université PSL, Sorbonne Université, CNRS, 92190 Meudon, France\\
 Sorbonne Universit\'e, CNRS, UMR 7095, Institut d'Astrophysique de Paris, 98 bis bd Arago, 75014 Paris, France
}

\date{\today}
\begin{abstract}
We introduce the $r$-largest order statistics ($r$-LOS) framework for galaxy clusters, a generalization of the standard extreme value statistics (EVS) approach. While conventional EVS focuses on the probability distribution of the most massive cluster within a given volume, the $r$-LOS framework exploits the additional statistical information encoded in the joint distribution of the $r$ most massive clusters, ranked in descending order of mass. We develop the corresponding mathematical formalism, discuss its asymptotic properties, and derive a set of relations relevant to galaxy cluster cosmology. In particular, we demonstrate that the high-mass end of the halo mass function (HMF) can be reconstructed from the generalized extreme value (GEV) parameters inferred using an extension of the block maxima method that incorporates the $r$ most massive clusters in each volume. This provides a means of recovering the high-mass tail of the HMF using only a small subsample of the halo population. We validate these relations using both comoving and light-cone halo catalogues from numerical simulations. Finally, we investigate the sensitivity of the $r$-LOS framework to primordial non-Gaussianity, focusing on its impact on the statistics of massive clusters. 
\end{abstract}
\maketitle
\section{Introduction}
Galaxy clusters are gravitationally bound systems residing in massive dark matter halos and represent the endpoint of the hierarchical, bottom-up process of cosmic structure formation \citep{allen2011,kravtsov2012}. Their properties therefore encode valuable information about both cosmology and astrophysical processes. In particular, the abundance of the most massive clusters probes the high-mass end of the halo mass function (HMF), which describes the number density of halos as a function of mass and redshift. In this regime, the HMF is exponentially suppressed, reflecting the extreme rarity of these objects. Their abundance is consequently highly sensitive to the underlying cosmological model and provides a powerful probe of the growth of cosmic structure.

This sensitivity originates from the fact that the most massive clusters form the highest peaks of the primordial matter density field. Their abundance therefore depends strongly on the amplitude of matter density fluctuations, commonly parametrized by $\sigma_8$, and the cosmic mean matter density, $\Omega_{\rm m}$. Furthermore, it can retain signatures of departures from the standard cosmological paradigm. In particular, several studies have demonstrated the potential of massive galaxy clusters to constrain primordial non-Gaussianity (PNG), which alters the statistical properties of rare density peaks in the initial conditions of structure formation (e.g., \cite{1988ApJ...330..535L,matarrese2000,verde2000,loverde2008,chongchitnan_primordial_2012,Cay_n_2011,Harrison_2012,Hotchkiss_2011}).

Extreme Value Statistics (EVS) provides a natural statistical framework for quantifying the extremeness of the most massive galaxy clusters and addresses the fundamental question: \textit{Is the most massive observed galaxy cluster consistent with a given cosmological model?} To date, EVS has been primarily employed to derive cosmological exclusion curves in the cluster mass-redshift plane, thereby assessing whether individual clusters are compatible with the predictions of a given cosmological model within the survey volume (see, e.g., \cite{Mortonson_2011,Hotchkiss_2011,Harrison_2012,waizmann_modelling_2012,waizmann_order_2013,Holz_2012}). These analyses have generally found no significant tension with the predictions of the standard $\Lambda$ Cold Dark Matter ($\Lambda$CDM) model. More recently, EVS has also been used to constrain cosmological parameters, such as $\sigma_8$ \citep{Busillo_2023}, and, when combined with cosmic void statistics, to help break the $\Omega_{\rm m}-\sigma_8$ degeneracy \citep{Sahlen:2015wpc}.

The advent of wide-area galaxy cluster surveys, such as \textit{Planck} \cite{2016A&A...594A..27P}, \textit{eROSITA} \cite{2024A&A...685A.106B}, and \textit{Euclid} \cite{2025_euclid_mission}, opens new opportunities to move beyond consistency tests based on individual objects and exploit the full statistical distribution of extreme cluster masses. In this regard, \citet{waizmann_novel_2011} demonstrated that comparing the Generalized Extreme Value (GEV) distribution, which describes the asymptotic behavior of extreme-value statistics, with the exact EVS distribution provides a significantly more stringent test of cosmological models.

In this work, we extend the standard EVS framework beyond the most massive cluster in a given sample by considering the statistical distribution of the $k$-th most massive objects. These higher-order statistics provide additional information on the high-mass end of the halo population (see \cite{waizmann_novel_2011}), as the masses of ranked objects are not statistically independent but are instead correlated through their ordering. Within the asymptotic EVS formalism, the distributions of all order statistics are fully determined by the same set of GEV parameters that characterize the distribution of maxima. This property allows us to generalize the inference procedure from the single most massive clusters to the joint distribution of the $r$-most massive systems in a given volume, referred to as the $r$-Largest Order Statistics ($r$-LOS) approach \citep{coles_introduction_2001}. By incorporating information from multiple extreme objects simultaneously, this approach provides a more comprehensive statistical characterization of the high-mass end of the halo mass function (HMF), thus providing more stringent cosmological consistency tests.

Originally developed to characterize extreme phenomena in hydrology, meteorology, and climate science (e.g., \cite{smith1986,tawn1988}), the $r$-LOS approach enables a more efficient use of available data by exploiting the joint statistical information carried by multiple extreme events. This is particularly advantageous in the cosmological context, where the scarcity of very massive clusters limits the statistical power of analyses based solely on the most massive object. Here, we apply this methodology to reconstruct the full set of order-statistic distributions from cluster samples, thereby extending the traditional EVS question to: \textit{Are the $r$-most massive observed clusters consistent with a given cosmological model?} This approach is further motivated by the direct connection between the GEV parameters and the expected number density of halos above a given mass threshold in the Poisson limit, which links the statistical properties of extreme clusters to the high-mass end of the HMF.
This relation establishes a novel connection between the statistics of rank-ordered massive clusters and the high-mass tail of the HMF, which we exploit here to investigate the sensitivity of this approach to primordial non-Gaussianity.

Galaxy cluster masses can be estimated using a variety of observational proxies, depending on the available data, including X-ray luminosity, the Sunyaev–Zel'dovich signal, weak gravitational lensing, and optical richness. However, these mass estimates are subject to systematic uncertainties and calibration biases (e.g., \cite{mantz2010,pratt2019}), which can mimic or mask departures from the predictions of the standard $\Lambda$CDM model. Since the primary aim of this work is to introduce and validate the $r$-LOS framework, we deliberately restrict our analysis to mock catalogues, for which the true halo masses are known. This allows us to isolate the statistical performance of the method from the additional uncertainties associated with observational mass estimation.

The paper is organized as follows. Section~\ref{sec:LOS} introduces the LOS statistical formalism, followed in Section~\ref{sec:LOS_phe} by a set of phenomenological applications. Section~\ref{nbody_data} describes the $N$-body simulation datasets used to validate the method in both comoving and redshift space. The results of these validation analyses are presented in Section~\ref{result_hmf_recon}. Section~\ref{sec:PNG} then investigates the impact of local primordial non-Gaussianity on the order statistics and assesses the constraining power of the proposed approach. Finally, Section~\ref{conclu} summarizes our main findings and conclusions.

\section{Largest order statistics}\label{sec:LOS}
Here, we introduce the largest order statistics for rank-ordered massive dark matter halos that host galaxy clusters, both in the finite and large sample limits. We refer interested readers to the textbook by S. \citet{coles_introduction_2001} and that by  \citet{Arnold2008OrderStatistics} for an exhaustive mathematical review of the subject. 

To clarify the notation, let us consider a sample of $N$ halos ranked in ascending mass order (from least to most massive):
\begin{equation}\label{eq:ranked_order}
M^{(N)} < \ldots < 
\underbrace{M^{(r)} < \ldots < M^{(k)} < \ldots < M^{(1)}}_{r-\text{largest elements}}.
\end{equation}
The standard EVS approach focuses on the statistics of the largest halo mass, $M^{(1)}$ in Eq.~(\ref{eq:ranked_order}), both in the finite and in the large-$N$ limit (i.e. $N\to +\infty$). This is with the intent of establishing a link between the HMF and the probability distribution of the most massive halo. Our goal is to extend the formalism to higher-order ranked halos by deriving: a) the statistics of the $k$-th largest halo (single order), corresponding to evaluating the probability distribution of $M^{(k)}$ in Eq.~(\ref{eq:ranked_order}); b) the statistics of the $r$-largest halos (multiple order), i.e. the joint distribution of $\{M^{(1)},\cdots,M^{(r)}\}$ in Eq.~(\ref{eq:ranked_order}).

In the following, we present a set of fundamental mathematical relations that establish a connection between these rank-ordered statistics and the abundance of halos, which form the basis for the inference  method developed in this work.

\subsection{Exact single order model formulation and large-$N$ limit}
Let us consider a halo of mass $M$. We can think of the value of the halo mass as the realization of a random variable drawn from a probability distribution function (PDF) specified by the HMF, $dn/dM$. This is the comoving number density of halos with mass in the infinitesimal interval $[M,M+dM]$. In such a case, the associated cumulative distribution function (CDF), i.e. the probability of finding a halo with mass smaller than $m$, is given by the ratio of the number density of halos with mass in the range $[m_{\rm min},m]$ divided by the total number density of halos, $n_{\rm tot}$, where $m_{\rm min}$ is a lower-bound corresponding to a minimal observable mass. This reads as:
\begin{equation}\label{fm}
F(m)=\frac{1}{n_{\rm tot}}\int_{m_{\rm min}}^m \frac{dn}{dM}\,dM,
\end{equation}
where 
\begin{equation}\label{ntot}
n_{\rm tot}=\int_{m_{\rm min}}^{+\infty} \frac{dn}{dM}\,dM.
\end{equation}

Consider now a comoving volume $V$ containing $N=n_{\rm tot}V$ independent and identically distributed halos with masses $\{M_1,\ldots,M_N\}$, where each mass is drawn from $F$. Let $M_\textrm{max}\equiv M^{(1)}=\max\{M_1,\ldots,M_N\}$ be the maximum of the $N$ halo masses, the probability that the most massive halo has a mass smaller than $m$ is given by the CDF:
\begin{align}
\Phi_N(M_\textrm{max}\le m) &= \Phi_N(M_1\le m, \ldots, M_N \le m)\nonumber\\
&= F(M_1\le m) \cdots F(M_N\le m)\nonumber \\
&= F^N(m),\label{pmax}
\end{align}
which is the exact EVS distribution of the most massive halo in a sample with finite number of elements \cite{harrison_exact_2011}. Notice that this derivation holds as long as the halo masses can be treated as independent random variables. This is the case only on sufficiently large scales ($L \gtrsim 100\,{\rm Mpc}\,h^{-1}$), where massive halos are spatially uncorrelated. On smaller scales, modifications induced by spatial correlations of the halos need to be taken into account (see \cite{davis_most_2011}).

The above result can be generalized to obtain the CDF of the $k$-th most massive halo of mass $M^{(k)}_{\rm max}\equiv M^{(k)}$ in the volume $V$ (see e.g. \cite{waizmann_order_2013,casella2002statistical} for a detailed derivation):
\begin{equation}\label{exact_rank}
\Phi_{N}^{(k)}(m)=\sum_{l=0}^{k-1}{\binom{N}{l}}\left[1-F(m)\right]^l F(m)^{N-l},
\end{equation}
This is the exact LOS distribution of the top-$k$ ranked halo in a finite sample. Notice that in the case $k=1$, Eq.~(\ref{exact_rank}) reduces to the EVS result, which is therefore a particular case of the more general LOS framework.  
In the large-$N$ limit, the CDF of the $k$-th most massive halo Eq.~(\ref{exact_rank}) reads as:
\begin{equation}
\Phi^{(k)}(m)\equiv\lim_{N\to\infty}\Phi_{N}^{(k)}(m)=e^{-\mathcal{N}(m)}\sum_{l=0}^{k-1} \frac{\mathcal{N}(m)^l}{l!}.
\end{equation}
where $\mathcal{N}(m)\equiv n(>m)V$ is the number of halos with mass greater than $m$. We refer the reader to Appendix~\ref{appendix_asym_k_gev} for a derivation. Note that the right-hand side of the above equation is the regularized upper incomplete gamma function, $\mathcal{Q}(k,\mathcal{N}(m))$\footnote{The regularized upper incomplete gamma function is given by:
\begin{equation}
    \mathcal{Q}(k,\mathcal{N}(m)) = \frac{1}{(k-1)!}\int_{\mathcal{N}(m)}^{\infty}t^{k-1}e^{-t}dt\nonumber
\end{equation}}. Hence, the CDF of the $k$-th most massive halo is given by
\begin{equation}\label{ranked_HMF}
    \Phi^{(k)}(m) = \mathcal{Q}(k,\mathcal{N}(m)),
\end{equation}
and the associated PDF can be obtained by taking the derivative with respect to the halo mass. This reads as
\begin{equation}\label{exact_distrib}
    \phi^{(k)}(m) = -\frac{\mathcal{N}(m)^{k-1}e^{-\mathcal{N}(m)}}{(k-1)!}\mathcal{N}'(m),
\end{equation}
where
\begin{equation}\label{der_N}
    \mathcal{N}'(m) \equiv \frac{d\mathcal{N}}{dm} = -V\frac{dn}{dm}.
\end{equation}
is the negative expected number of halos with mass in the infinitesimal interval $[m,m+dm]$ in the volume $V$. 
Notice that the formulae above can be easily generalized to the case of a survey volume defined by a solid angle $\Delta\Omega$ and covering a redshift interval of size $\Delta{z}$. In such a case, the number of halos is given by
\begin{equation}\label{nc_survey}
\mathcal{N}(m,z)=\Delta\Omega\int_{z-\frac{\Delta z}{2}}^{ z+\frac{\Delta z}{2}}\int_m^{+\infty}\frac{{\rm d}^2V}{{\rm d}\Omega {\rm d}z'} \frac{{\rm d}n}{{\rm d}M}\,{\rm d}M{\rm d}z',
\end{equation}
where ${\rm d}^2V/{\rm d}\Omega {\rm d}z'$ is the infinitesimal comoving volume element.

\subsection{Void probability distribution} 
A different approach to compute the statistics of the most massive halo in a given volume $V$ is to consider the void probability distribution of a Poisson process, usually denoted as $P_0$, which reads as: 
\begin{equation}\label{void_distrib}
    P_0(m) = e^{-\mathcal{N}(m)},
\end{equation}
that is the probability that in a given volume $V$ there are no halos with mass greater than $m$ \cite{davis_most_2011,colombi_extreme_2011,chongchitnan_primordial_2012}. 

By comparing Eq.~(\ref{void_distrib}) with Eq.~(\ref{ranked_HMF}) in the case $k=1$, we find that $P_0(m)\equiv \mathcal{Q}(1,\mathcal{N}(m))$. This relation implies that the void distribution corresponds to the large-$N$ limit of the exact EVS formulation for the most massive object in a given volume $V$. Consequently, Eq.~(\ref{ranked_HMF}) can be interpreted as a generalization of the void distribution to the case of the top-$k$ rank-ordered halo mass. Hence, it gives the probability that at most $k-1$ halos have mass greater than $m$ within a given volume $V$. Furthermore, the void distribution has been used to account for the impact of spatial correlations among massive halos within the EVS framework. This correspondence provides a natural way to incorporate such effects into Eq.~(\ref{ranked_HMF}). In particular, following \cite{chongchitnan_primordial_2012}, the contribution of spatial correlations can be included through a shift in the argument of Eq.~(\ref{ranked_HMF}) of the form
\begin{equation}
\mathcal{N}(m) \rightarrow \mathcal{N}(m) - \frac{1}{2}\left[\mathcal{N}(m)\beta_L\sigma(L)\right]^2 ,
\end{equation}
where $\beta_L$ denotes the effective halo bias on the scale $L$, and $\sigma(L)$ is the root mean square of the linear density field smoothed on the same scale.

\subsection{Asymptotic model formulation}
A key result of the EVS formalism is that, even when the cumulative distribution function of $M_{\rm max}$ is not explicitly known, its distribution converges to a universal asymptotic form in the large-$N$ limit. This limiting distribution is the so-called Generalized Extreme Value distribution, which describes the statistics of block\footnote{A block refers to a given volume $V$ and hereafter we will use these terms interchangeably.} maxima.

Specifically, when $M_{\rm max}$ is appropriately normalized by two constants $a_N$ and $b_N>0$, the rescaled variable $\bar{M}_{\rm max} = (M_{\rm max} - a_N)/b_N$ has, in the large-$N$ limit, a cumulative distribution function that converges to the following form \cite{Fisher_Tippett_1928,gnedenko}: 
\begin{equation}\label{gev_blockmaxima}
G(m)\equiv\lim_{N\to\infty}\Phi_N(\Bar{M}_\textrm{max}\le m) = e^{-\tau(m)}.
\end{equation}
where
\begin{equation}
    \tau(m) = 
    \begin{cases}
        \left[1 + \xi\left(\dfrac{m-\mu}{\sigma}\right) \right]^{-1/\xi} & \text{if } \xi \neq 0\\[0.2em]
        \exp\left\{- \left(\dfrac{m-\mu}{\sigma}\right)\right\} & \text{if } \xi = 0
    \end{cases}
\end{equation}
depends on three parameters: the location parameter $\mu$, that is generally defined in the interval $-\infty<\mu<+\infty$, however in the case of halo masses this is restricted to $\mu\geq0$; the scale parameter $\sigma$, that is positive definite, $\sigma>0$; the shape parameter $\xi$, where the parameters satisfy the condition $1+\xi(m-\mu)/\sigma > 0$. Note that, since the normalization constants $a_N$ and $b_N$ can be absorbed into $\mu$ and $\sigma$, the asymptotic limit of the CDF of $M_{\rm max}$ is also given by Eq.~(\ref{gev_blockmaxima}), where $\mu$ and $\sigma$ become volume dependent.

The location parameter sets the central tendency of the GEV, $\sigma$ determines the level of scatter around $\mu$, while $\xi$ specifies the type of distribution. In particular, in the case of $\xi = 0$, we have the Gumbel distribution (Type I), for $\xi > 0$, we have the Fréchet distribution (Type II), while $\xi < 0$ corresponds to the Weibull distribution (Type III). These three families of distributions differ primarily in the behavior at large $m$ values. For instance, in the Gumbel case ($\xi=0$), the tail of the distribution is characterized by an exponential decay and the absence of a finite upper bound. Similarly, the Fr\'echet distribution ($\xi>0$) has no finite upper limit, but exhibits a heavy tail with power-law behavior, implying that extremely large events may occur with a non-negligible probability. In contrast, the Weibull distribution ($\xi<0$) possesses a finite upper endpoint, beyond which extreme events cannot occur, and therefore describes physically bounded extreme processes. For this reason, the statistical occurrence and physical interpretation of extreme events depends crucially on the value of the shape parameter $\xi$.

As in the exact EVS formulation, the GEV result can also be generalized to the LOS of the top-$k$ largest halo with mass $M^{(k)}$. In particular, assuming the Poisson limit, and adopting the same normalization used for the block maxima, i.e. $\bar{M}^{(k)} = (M^{(k)}-a_N)/b_N$, the cumulative distribution function of $\bar{M}^{(k)}$ converges to the top-$k$ GEV distribution (see \cite{coles_introduction_2001}, Chapter 7):
\begin{eqnarray}\label{kgev}
G_k(m) & = & \lim_{N\to\infty}\Phi_N^{(k)}(\bar{M}^{(k)}\le m) \nonumber\\
& = & \exp\left\{-\tau(m)\right\}\sum_{s=0}^{k-1}\frac{\tau(m)^s}{s!} \nonumber \\
& = & \mathcal{Q}(k,\tau(m)).
\end{eqnarray}

Again, the normalization constants $a_N$ and $b_N$ can be absorbed into the scale and location parameters. Consequently, the cumulative distribution function of $M^{(k)}$ is also described by Eq.~(\ref{kgev}). Finally, by taking the derivative of Eq.~(\ref{kgev}) with respect to the halo mass, we obtain the associated PDF:
\begin{equation}\label{gev_fit}
    g_k(m) = \frac{\tau(m)^{k+\xi}}{\sigma(k-1)!}e^{-\tau(m)}.
\end{equation}

\section{Phenomenology}\label{sec:LOS_phe}

In the previous section, we introduced the LOS formalism and derived the exact distribution of the $k$-th most massive cluster, establishing its connection to the asymptotic GEV description. Building on these results, we now explore several phenomenological implications, focusing on the statistical characterization of massive halos and the potential of LOS statistics for cosmological applications.

\subsection{HMF reconstruction with GEV parameters}\label{HMF_and_GEV}

\subsubsection{HMF and GEV equivalence in the large sample limit}
An important phenomenological implication of the LOS framework is that the precise estimation of GEV parameters enables the reconstruction of the high-mass end of the HMF. Indeed, let
us consider the exact CDF of the $k$-th most massive cluster in the large-$N$ limit, as given by Eq.~(\ref{ranked_HMF}), and compare it with the asymptotic form of the top-$k$ ranked GEV distribution in Eq.~(\ref{kgev}). This comparison yields
\begin{equation}\label{number_counts}
     \mathcal{N}(m) =\tau(m),
\end{equation}
implying that in a given volume $V$ the number of clusters with mass greater than $m$ at the high-mass end is well approximated by the function $\tau(m)$ specified by GEV parameters $\Theta_{\rm GEV}=(\mu,\sigma,\xi)$. Then, taking the derivative of both sides of Eq.~(\ref{number_counts}), we obtain
\begin{align}\label{equivalence_HMF}
\frac{dn}{dm} &= \frac{-1}{V}\frac{d\tau}{dm}= \frac{\tau(m)^{\xi+1}}{\sigma V},\nonumber\\
&=
    \begin{cases}
        \dfrac{1}{\sigma V}\left(1+\xi\left(\dfrac{m-\mu}{\sigma}\right) \right)^{-1/\xi -1} & \text{if } \xi \neq 0\\[0.7em]
        \dfrac{1}{\sigma V}\exp\left\{- \left(\dfrac{m-\mu}{\sigma}\right)\right\} & \text{if } \xi = 0
    \end{cases}
\end{align}

This suggests that it should be possible to reconstruct the high-mass end of the HMF from measurements of the GEV parameters inferred from mass estimates of a subsample of massive rank-ordered clusters, rather than sampling the full halo population.

\begin{figure*}
    \centering
    \includegraphics[width=1
    \linewidth]{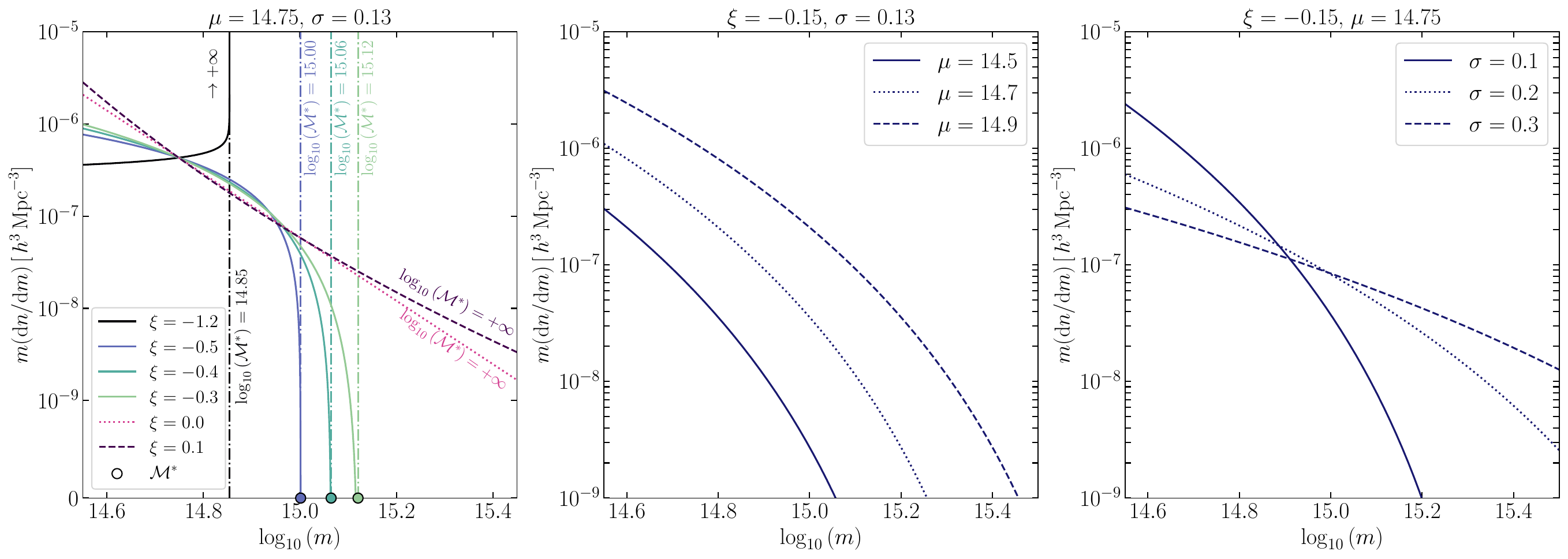}
    \caption{Impact of GEV parameters on the high-mass end of the HMF for a volume $V = (200\,\rm Mpc\,h^{-1})^{3}$. The left, middle, and right panels show the effect of varying $\xi$, $\mu$, and $\sigma$ respectively, with the remaining parameters held fixed at their fiducial value. On the left panel, the solid lines refer to the Weibull distributions ($\xi<0$), the dotted line to the Gumbel distribution ($\xi=0$), and the dashed line to the Fréchet family of distributions ($\xi>0$). The vertical dash-dotted lines correspond to the limiting mass  $\mathcal{M}^*$ given by Eq.~(\ref{universal_barrier}), that is the intrinsic upper bound of the Weibull attraction domain represented by black circles. The middle panel shows the impact of three different values of $\mu$ for fixed $\xi$ and $\sigma$, whereas the right panel shows three different $\sigma$ for fixed values of $\xi$ and $\mu$.}
    \label{fig:GEVtoHMF}
\end{figure*}

\subsection{Halo mass upper limit}
\subsubsection{Physicality of the HMF-GEV relation}

The validity of Eq.~(\ref{equivalence_HMF}) establishes a direct connection between the GEV parameters and the high-mass tail of the HMF. To illustrate this correspondence, Fig.~\ref{fig:GEVtoHMF} shows how the high-mass end of the HMF varies with different combinations of GEV parameters, assuming a fiducial volume of $V = (200\,\mathrm{Mpc}\,h^{-1})^3$ and expressing the mass in $\log_{10}(m)$. In the left panel, we vary the shape parameter $\xi$ while keeping $\mu$ and $\sigma$ fixed to $\mu=14.75$ and $\sigma=0.13$. In the middle and right panels, we fix $\xi=-0.15$ and vary $\mu$ and $\sigma$, respectively. The middle panel shows that, at fixed $\sigma=0.13$, increasing values of $\mu$ shift the HMF toward higher masses. Conversely, the right panel illustrates that, at fixed $\mu=14.75$, increasing $\sigma$ leads to a progressive flattening of the HMF slope.

We next investigate the role of the shape parameter $\xi$. As illustrated in the left panel, changing $\xi$ from negative to positive values leads to substantial changes in the asymptotic behavior of the HMF. For $\xi<0$, the high-mass tail is characterized by the Weibull distribution; for $\xi=0$, it corresponds to the Gumbel distribution; and for $\xi>0$, to a Fr\'echet distribution. This demonstrates that the sign of $\xi$ governs the asymptotic behavior of the HMF. For $\xi \geq 0$, the HMF decreases asymptotically toward zero, as illustrated by the purple dashed and magenta dotted curves in the left panel of Fig.~\ref{fig:GEVtoHMF}. By contrast, for $\xi < 0$, the distribution exhibits a finite upper mass limit, $\mathcal{M}^*$. In particular, for $\xi=-1.2$, the HMF diverges at the high-mass end as the halo mass approaches this upper bound, exhibiting a vertical asymptote at $m=\mathcal{M}^*$. The limiting case $\xi=-1$ (not shown in Fig.~\ref{fig:GEVtoHMF}) is also noteworthy: the HMF becomes independent of halo mass, remaining constant at $1/(V\sigma)$ for $m<\mathcal{M}^*$. Such a behavior is clearly unphysical, as the abundance of halos is expected to decrease monotonically with increasing mass. We therefore restrict physically meaningful descriptions of the HMF to $\xi>-1$. For intermediate values of the shape parameter, $-1 < \xi < 0$, the HMF approaches a finite upper mass endpoint at larger masses given by
\begin{equation}\label{universal_barrier}
\mathcal{M}^* = 10^{\mu - \frac{\sigma}{\xi}},
\end{equation}
which diverges in the limit $\xi \to 0^-$. Consequently, a GEV description of the high-mass tail with $-1 < \xi < 0$ implies a strict upper bound on the halo mass distribution. This corresponds to the Weibull domain of attraction, characterized by a compact support: the probability of finding a halo with mass exceeding $\mathcal{M}^*$ is identically zero.

We now turn to the physical implications of the different domains of attraction. Both the Fréchet and Gumbel distributions have unbounded support, implying a non-zero probability of finding virialized structures with arbitrarily large masses. Such a prediction is, however, physically unrealistic in a cosmological context. In particular, it would allow for a non-zero probability of finding halos more massive than the total matter content of the observable Universe, an outcome that is clearly unphysical and should instead have strictly zero probability.

In the context of $\Lambda$CDM cosmology, several physical considerations suggest that the HMF may belong to the Weibull domain of attraction. First, the finite matter content of the observable Universe imposes a natural upper bound on the mass that can be assembled into a single gravitationally bound structure. Second, the hierarchical nature of CDM structure formation, whereby small-scale structures collapse first and subsequently merge to form increasingly massive halos, provides a further limitation on the growth of extreme-mass objects. In particular, the finite time elapsed between the formation of the first low-mass halos and the present epoch constrains the extent of their subsequent mass assembly. This limitation is further reinforced by the onset of dark energy domination, which suppresses the growth of matter perturbations and progressively reduces the efficiency of both halo mergers and mass accretion.

Taken together, these considerations point to the existence of an effective upper bound on halo masses, arising from the finite matter content of the observable Universe and the limited growth of structures in a hierarchical cosmological scenario. Accordingly, the HMF is expected to be in the Weibull domain of attraction, corresponding to $-1 < \xi < 0$, with a finite upper mass endpoint given by Eq.~(\ref{universal_barrier}).

\subsubsection{Observational implications}
\begin{figure}[t]
    \centering
    \includegraphics[width=0.9\linewidth]{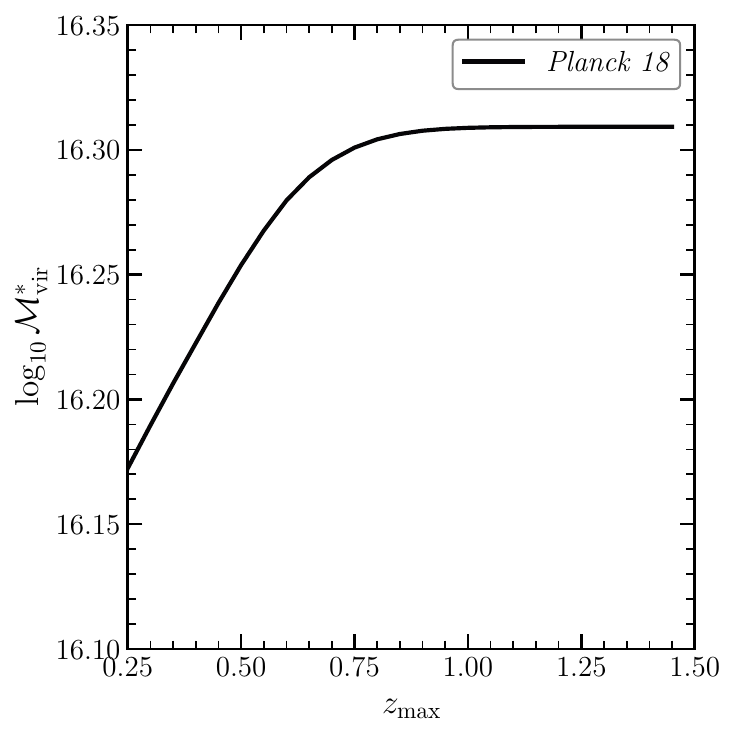}
    \caption{The upper mass limit as a function of $z_{\rm max}$ for a full-sky survey. The upper-mass limit reaches an asymptotic value at $z_{\rm max}\simeq 1$ corresponding to $\mathcal{M}^*_{\rm vir} \simeq 10^{16.3}\,{\it M}_\odot h^{-1}$.}
    \label{fig:mstar_allsky}
\end{figure}

The existence of an upper mass limit for massive halos can be exploited to perform cosmological consistency tests using rank-ordered galaxy cluster samples. To see this, let us consider Eq.~(\ref{universal_barrier}). Following the approach presented in \cite{davis_most_2011}, the GEV parameters can be expressed in terms of quantities that depend solely on the assumed HMF. Specifically, these relations are obtained by performing a Taylor expansion of both the GEV distribution and the void probability distribution around the peak value, $m_0$. This yields (we refer the reader to Appendix~\ref{appendix_peak} for a detailed derivation)
\begin{align}
     \xi &= \mathcal{N}_0-1,\label{HMF_to_xi}\\
     \sigma &= \frac{-(1+\xi)^{1+\xi}}{\mathcal{N}'({m_0})}, \label{HMF_to_sigma}\\
     \mu &= m_0- \frac{\sigma}{\xi}\left[(1+\xi)^{-\xi}-1 \right],\label{HMF_to_mu}  
\end{align}
where $\mathcal{N}_0 \equiv \mathcal{N}(m_0)$ denotes the number of clusters with mass $m>m_0$, and $\mathcal{N}'$ is defined in Eq.~(\ref{der_N}). The value of $m_0$ can be determined by solving the algebraic equation: 
\begin{equation}
    \mathcal{I}'(m_0)+\Delta\Omega\,\mathcal{I}^2(m_0)=0,
\end{equation}
where 
\begin{equation}
    \mathcal{I}(m_0) =\int_{z_{\rm min}}^{z_{\rm max}}\frac{{\rm d}n}{{\rm d}m}\left(m_0,z\right)\frac{{\rm d}^2V}{{\rm d}z{\rm d}\Omega}\left(z\right)dz, 
\end{equation}
$\mathcal{I}'(m_0)={\rm d}\mathcal{I}/{\rm d}m\vert_{m_0}$ and $\Delta\Omega$ is the galaxy cluster survey area.
Combining Eqs.~(\ref{HMF_to_xi})--(\ref{HMF_to_mu}) with Eq.~(\ref{universal_barrier}), the upper mass limit $\mathcal{M}^*$ can be expressed as
\begin{equation}\label{analytical_mstar}
\mathcal{M}^* = m_0 + \frac{\mathcal{N}_0}{\mathcal{N}_0 - 1}\,\frac{1}{\mathcal{N}'(m_0)}.
\end{equation}
Assuming an analytical form for the HMF, \ref{analytical_mstar} enables the prediction of $\mathcal{M}^*$ at any redshift and for any cosmological model. As an illustration, Fig.~\ref{fig:mstar_allsky} shows the redshift evolution of the limiting mass $\mathcal{M}^*$, defined at the virial overdensity, for a reference $\Lambda$CDM cosmology best-fit to {\it Planck} 2018 data \citep{2020A&A...641A...6P} (P18, hereafter). The limiting mass is computed using the \textit{Euclid}-HMF~\cite{Castro2023}, assuming a full-sky survey covering a solid angle of $4\pi\,\rm sr$, with $z_{\rm min}=0.0$ and $z_{\rm max}\in[0.25,1.50]$. We can see that the limiting mass increases monotonically as function of $z_{\rm max}$, eventually approaching an asymptotic value of $\mathcal{M}^* \simeq 10^{16.3}\,{\it M}_\odot h^{-1}$. This plateau is reached at approximately $z_{\rm max}\simeq 1.0$. This behavior reflects the fact that the most massive gravitationally bound structures require sufficient cosmic time to form and assemble, making their presence increasingly unlikely at higher redshifts. We therefore expect the most massive clusters to reside predominantly within the redshift interval $z\in[0,1.0]$. Consequently, if the masses of extreme clusters in the observable Universe are drawn from the Weibull distribution for the $\Lambda$CDM scenario, the probability of finding a cluster with a mass exceeding $\mathcal{M}^* \simeq 10^{16.3} {\it M}_\odot h^{-1}$ vanishes. This upper mass limit is nearly an order of magnitude larger than the estimated virial mass of El Gordo (ACT-CL J0102$-$4915, $z=0.87$), the most massive galaxy clusters known to date \citep{Diego2023}.

A few caveats are worth emphasizing. First, the actual \textit{Euclid}-HMF adopted here to compute the GEV parameters does not exhibit a strict upper mass limit, but rather an exponential cut-off that approaches zero only asymptotically. The negative value of the shape parameter, $\xi<0$, therefore arises from the finite-volume effect and should not be interpreted as evidence for an intrinsic upper bound in the underlying HMF. Consequently, the resulting value of $\mathcal{M}^*$ represents an effective, volume-dependent upper mass endpoint, rather than a universal physical limit. Moreover, its value depends sensitively on the adopted parametrization of the multiplicity function, which was neither derived nor calibrated within the EVS framework. Thus, $\mathcal{M}^*$ should be regarded as an effective upper limit conditional on both the survey volume and the adopted semi-analytical HMF. Nevertheless, if the HMF of the actual Universe does possess a finite upper mass limit, a possibility that remains to be established mathematically, this limit would be given by \refeq{analytical_mstar}.

\subsection{GEV parameter estimation and r-LOS}\label{GEVparams_from_rLOS}
Let us now discuss how to accurately estimate the GEV parameters. Several methods have been developed for this purpose (see e.g. \cite{coles_introduction_2001}). For example, in the peaks-over-threshold (POT) approach, extreme events are identified as the values that surpass a chosen threshold (see \citep{waizmann_modelling_2012} for an application to galaxy cluster observations).
An alternative approach is to estimate the distribution's moments from the corresponding population moments \citep{Hosking_Wallis_1997}. Specifically, one equates the analytical expressions of the moments (that depend on the GEV parameters) to the numerical values obtained from the population estimates (i.e., computed from a data sample), and then solves these relations to estimate the GEV parameters. By contrast, the most widely used approach is the block maxima method. It consists of partitioning the parent dataset into $N_V$ blocks of fixed size, extracting the most extreme observation from each block, and fitting a GEV distribution to the resulting sample of maxima to infer its parameters.

In the context of galaxy clusters, applying the block maxima method requires partitioning the dataset into sufficiently large blocks to ensure that spatial correlations between their respective maxima are negligible, while retaining enough blocks to adequately sample the EVS distribution. These competing requirements, however, limit the number of independent maxima available for the analysis, thereby increasing the statistical uncertainty and broadening the posterior distributions of the inferred GEV parameters.

Here, we show that extending the block maxima approach to the $r$-largest observations in each block significantly improves the constraints on the GEV parameters. 
As shown in \cite{coles_introduction_2001,Weissman1978}, in the asymptotic limit, the joint probability density of the $r$-largest order statistics can be expressed in the following form, provided that all components of the random vector $\mathbf{M}^{(r)} = \{M^{(1)},M^{(2)},\ldots,M^{(r)}\}$ share the same normalization constants:
\begin{equation}\label{joint_distribution}
f\left(\mathbf{m}^{(r)}|\Theta_{\rm GEV}\right) =  \exp \left\{-\tau\left(m^{(r)}\right) \right \}\prod_{k = 1}^{r} \frac{\tau\left(m^{(k)}\right)^{1+\xi}}{\sigma},
\end{equation}
where $\mathbf{m}^{(r)} = \{m^{(1)},\ldots,m^{(r)}\}$ denotes a realization of $\mathbf{M}^{(r)}$, corresponding to the masses of the $r$-most massive clusters, ordered such that $m^{(1)} \geq \cdots \geq m^{(r)}$. Here, $\Theta_{\rm GEV}=(\mu,\sigma,\xi)$ denotes the vector of GEV parameters. Provided that the blocks are sufficiently large, the vectors $\mathbf{m}_i^{(r)} = \{m_i^{(1)},\ldots,m_i^{(r)}\}$, with $i\in \{1,\ldots,N_V\}$, can be regarded as $N_V$ independent realizations of the random vector $\mathbf{M}^{(r)}$, where $m_i^{(k)}$ denotes the mass of the $k$-th most massive cluster in the $i$-th block. Under this assumption, the $r$-LOS likelihood follows directly from Eq.~(\ref{joint_distribution}) and can be written as
\begin{equation}\label{likelihood}
    \mathcal{L}(\mathbf{M}_{\rm obs}|\Theta_{\rm GEV}) = \prod_{i = 1}^{N_V}\exp \left\{-\tau\left(m^{(r)}_i\right) \right \} \prod_{k = 1}^{r} \frac{\tau\left(m^{(k)}_i\right)^{1+\xi}}{\sigma},
\end{equation}
where $\mathbf{M}_{\rm obs}=\{\mathbf{m}^{(r)}_1,\ldots,\mathbf{m}^{(r)}_{N_V}\}$ denotes the observed data vector. If a given block contains fewer than $r$ clusters, the likelihood can still be evaluated by replacing $r$ with $r_i$ in the corresponding contribution, where $r_i$ is the highest available rank in the $i$-th block.

Given the likelihood in Eq.~(\ref{likelihood}), the best-fit GEV parameters can be inferred from a sample of the $r$ most massive clusters across $N_V$ blocks by sampling the posterior probability distribution,
\begin{equation}\label{posterior_GEV}
\mathcal{P}(\Theta_{\rm GEV}|\mathbf{M}_{\rm obs})\propto \mathcal{L}(\mathbf{M}_{\rm obs}|\Theta_{\rm GEV})\Pi(\Theta_{\rm GEV}),
\end{equation}
where $\Pi(\Theta_{\rm GEV})$ denotes the prior probability distribution of the GEV parameters, which we take to be uninformative.

The standard block maxima approach corresponds to the case $r=1$, in which only the most massive object in each block is considered. Since the maxima extracted from the $N_V$ blocks are assumed to be statistically independent, the Cramér--Rao lower bound applies, implying that the variance of the estimated GEV parameters scales asymptotically as $\textrm{Var}(\Theta_{\rm GEV}) \simeq 1/N_V$. Extending the analysis to the $r$ most massive objects in each block allows us to exploit additional statistical information, as reflected in the likelihood of Eq.~(\ref{likelihood}). In this case, the variance retains the same scaling with the number of blocks but with a different proportionality coefficient, $\textrm{Var}(\Theta_{\rm GEV}) \simeq C(r)/N_V$, where $C(r)$ is a decreasing function of $r$. Thus, by incorporating multiple extreme objects, the $r$-LOS approach provides a more comprehensive sampling of the high-mass tail of the halo distribution, potentially improving the goodness-of-fit and reducing the statistical uncertainty and associated biases in the inferred GEV parameters. Furthermore, it allows us to use larger blocks, thereby reducing the impact of spatial correlations, while retaining information on the most massive systems, which are generally easier to detect and characterize observationally.

The choice of the rank $r$, however, involves a trade-off. Although increasing $r$ generally reduces the variance of the parameter estimates, choosing it too large may extend the sampled mass range beyond the asymptotic tail of the distribution, where the GEV approximation is no longer valid. This can introduce systematic biases that offset the gains in statistical precision. The optimal value of $r$ must therefore balance the reduction in variance against the requirement that the selected objects remain within the asymptotic regime. In practice, this can be assessed using techniques such as parametric bootstrap methods or by estimating the entropy difference between the $r$-LOS and $(r-1)$-LOS distributions, as proposed in \cite{bader_automated_2017}.

Overall, an unbiased estimate of the GEV parameters allows us to fully characterize the joint distribution of the $r$ most massive clusters. As we demonstrate below, this approach offers significantly greater sensitivity to cosmological parameters than the standard block maxima method described in \cite{waizmann_novel_2011}, providing a more powerful statistical framework for testing cosmological models.

\section{$N$-body Simulation Datasets}\label{nbody_data}
In this section, we describe the numerical simulation datasets used to validate the statistical relations derived in Section~\ref{sec:LOS}. To this end, we use the halo catalogue from a snapshot of the Uchuu simulation \citep{Ishiyama_2021} to perform the validation in comoving space, and the light-cone halo catalogue from the Flagship simulation \citep{EuclidSkyFlagship} to test the method in redshift space.

\subsection{Uchuu halo catalog and HMF}\label{sec:Uchuu}
We use the catalogs of halos at $z=0$ and $z=1$ from the large box of the Uchuu $N$-body simulation suite \cite{Ishiyama_2021} characterized by a $(2\,h^{-1}\,{\rm Gpc})^3$ volume with $12\,800^3$ particles, corresponding to a mass resolution of $m_p=3.3\times 10^8 M_{\odot}\,h^{-1}$. The simulation has been realized with the \texttt{GreeM} code \cite{2009PASJ...61.1319I,2012arXiv1211.4406I} for a flat $\Lambda$CDM model with cosmological parameters set to the best-fit \textit{Planck}-2015 cosmology \citep{Planck2016}: $\Omega_{\rm m} = 0.3089$, $\Omega_{\rm b} = 0.0486$, $h = 0.6774$, $n_{\rm s} = 0.9667$, $\sigma_8 = 0.8159$. Halos were detected using the \texttt{Rockstar} halo finder \cite{2013ApJ...762..109B} and are publicly available on CosmoHub\footnote{\url{https://cosmohub.pic.es/catalogs/316}} \cite{2017ehep.confE.488C,TALLADA2020100391}. In our analysis, we consider spherical overdensity halo masses defined within a radius enclosing a density $\Delta=500$ times the critical density of the Universe. 

To perform validation tests of the GEV--HMF relations using the $r$-LOS approach discussed in Section~\ref{sec:LOS_phe}, we compare the reconstructed HMF to a numerical estimate of the Uchuu-HMF, which we refer to as \textit{exact} HMF, given by
\begin{equation}\label{HMF_estimate_Uchuu}
\frac{{\rm d}n}{{\rm d}\log_{10}{M}}(M,z)= \frac{1}{V_{\rm tot}}\frac{N(M,z)}{\Delta\!\log_{10}{M}}\;,
\end{equation}
where $V_{\rm tot}$ is the simulation volume, $N(M,z)$ is the number of halos with mass $M$ in logarithmic mass bins of size $\Delta\!\log_{10}{M}$ at redshift $z$. We estimate the Poisson error in each bin as:
\begin{equation}\label{shot_noise}
    \sigma_{{\rm d}n/{\rm d}\log_{10}{M}} =\frac{1}{V_{\rm tot}}\frac{\sqrt{N(M,z)}}{\Delta\!\log_{10}{M}}\;.
\end{equation}
The size of the mass bins is set to guarantee that the highest mass bin contains at least one halo. 

\subsection{Flagship light-cone catalog simulation}\label{sec:flagship}
A more realistic assessment of the $r$-LOS formalism can be performed using the light-cone halo catalog from the Flagship simulation \cite{EuclidSkyFlagship}, designed to mimic the upcoming \textit{Euclid} DR1 galaxy cluster survey. The simulation consists of a volume of $(3\,h^{-1}\,\mathrm{Gpc})^3$ with $16\,000^3$ particles, corresponding to a particle mass of $m_{\rm p} = 10^9M_{\odot}\,h^{-1}$. It was carried out using the \texttt{PKDGRAV3} code \citep{2016asclsoft09016P}, assuming a $\Lambda$CDM cosmology with parameters set to the \textit{Euclid} fiducial values: $\Omega_{\rm m} = 0.319$, $\Omega_{\rm b} = 0.049$, $h = 0.67$, $n_{\rm s} = 0.96$, $\sigma_8 = 0.813$, $\Omega_{\rm r} = 5.509 \times 10^{-5}$, $\Omega_{\nu} = 0.00140343$, and $w = -1.0$.

As in the case of the Uchuu simulation, halo catalogs were constructed using the \texttt{Rockstar} halo finder. In this work, we make use of the WIDE light-cone dataset, corresponding to one octant of the sky with a solid angle $\Delta\Omega \simeq 5157\,\mathrm{deg}^2$, defined by the angular ranges $0 \leq \mathrm{RA} \leq 90^\circ$ and $0 \leq \mathrm{Dec} \leq 90^\circ$. The halo catalog is publicly available on CosmoHub\footnote{\url{https://cosmohub.pic.es/catalogs/352}} \citep{2017ehep.confE.488C,TALLADA2020100391}. Also in this case we consider halo masses defined at $500$ times the critical density of the universe.

\section{Numerical Tests}\label{result_hmf_recon}

We are now in a position to test the validity of Eq.~(\ref{equivalence_HMF}) and Eq.~(\ref{number_counts}) using the numerical simulation datasets introduced in the previous section. 

\subsection{Uchuu comoving space analysis}

\subsubsection{Enhanced GEV parameter inference with $r$-LOS}\label{GEV_inf}
We use the Uchuu halos to infer the parameters of the underlying GEV distribution. Specifically, we compare the results obtained with the $r$-LOS approach described in Section~\ref{GEVparams_from_rLOS} against the standard EVS estimation based on the block maxima method. The latter corresponds to considering only the most massive cluster ($r=1$) in each block, while the former extends the statistics to the $r$ ranked most massive systems. In the following, we work with the logarithmic cluster mass, $u = \log_{10}(m)$, rather than the mass itself.

Firstly, we divide the Uchuu simulation snapshot into $N_V=125$ blocks of volume $(400\,h^{-1}\,{\rm Mpc})^3$ each and we identify the ranked eight most massive halos. Then, we infer constraints on the $\Theta_{\rm GEV}$ parameters from the posterior distribution given in Eq.~(\ref{posterior_GEV}) in the case $r=1$ and $r=8$ respectively. To this purpose, we perform a  Markov Chain Monte Carlo (MCMC) sampling of the posterior using the \texttt{emcee} sampler \citep{Foreman_Mackey_2013}. 

We do not make any assumption on the attraction domain of the distribution of clusters, hence we assume flat priors on $\xi \sim U(-1,1)$. Similarly, we assume uniform priors on $\mu \sim  U(12,17)$ and $\sigma \sim U(0,3)$ with $1+\xi(u_i^{(j)}-\mu)/\sigma > 0$ for $j\in \{1,\ldots ,r\}$. In one of the $r$ ranked halos does not satisfy the latter condition, the likelihood is set to zero.

\begin{figure}[t]
    \centering
    \includegraphics[width=1.0\linewidth]{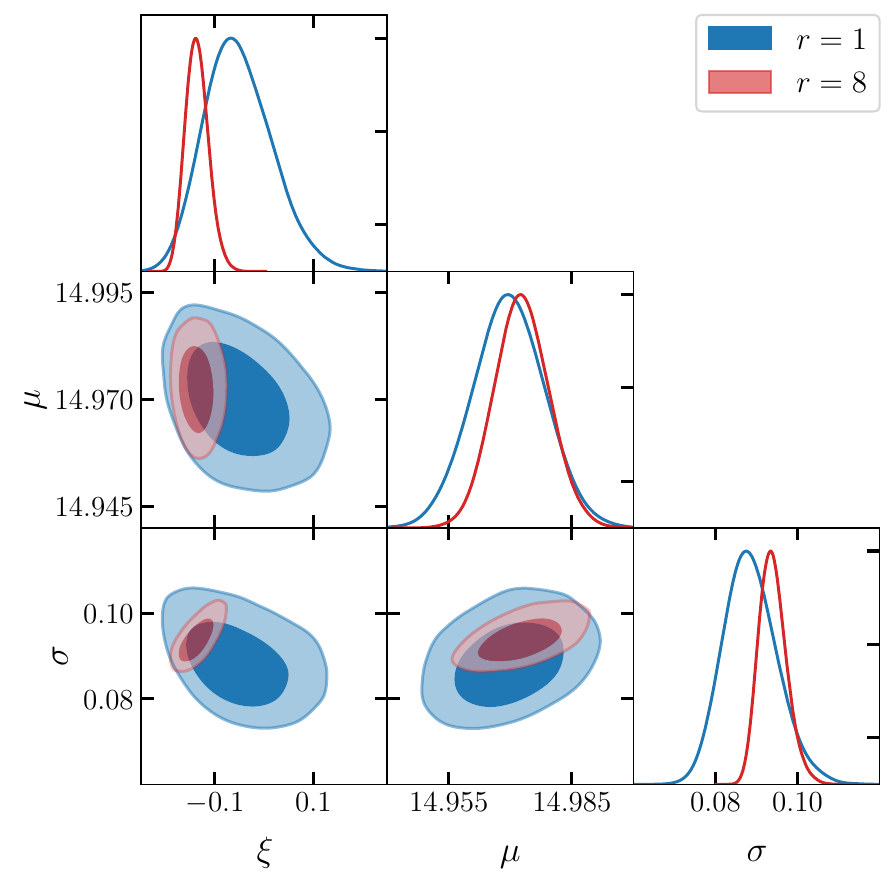}
    \caption{Triangular plot of the marginalized posterior distribution of the GEV parameters inferred from the Uchuu halos at $z=0$ in the EVS case ($r=1$) and $r$-LOS with $r=8$.}
    \label{fig:GEV_triangular_posterior}
\end{figure}

In Fig.~\ref{fig:GEV_triangular_posterior}, we show a triangular plot of the marginalized posteriors of the GEV parameters inferred for $r=1$ and $r=8$, respectively. As we can see, the use of the $r$-LOS approach results in narrower posterior distributions of the GEV parameters than the standard EVS method ($r=1$).
We further assess the performance of the two approaches by comparing the predicted top-$k$ GEV distribution, $G_k$, obtained from Eq.~(\ref{kgev}) using the inferred GEV parameters, with its empirical counterpart, $\hat{G}_k$. The empirical distribution is estimated from the sample of $k$-th most massive cluster masses across the $N_V$ blocks, sorted in ascending order as $m_1^{(k)} \leq m_2^{(k)} \leq \ldots \leq m_{N_V}^{(k)}$. 
It is then estimated using the plotting-position estimator:
\begin{equation}
    \hat{G}_k\left(m_i^{(k)}\right) = \frac{i}{N_V+1},
\end{equation}
where $i \in \{1, \ldots, N_V\}$.

\begin{figure}[ht]
    \centering
    \includegraphics[width=1.0\linewidth]{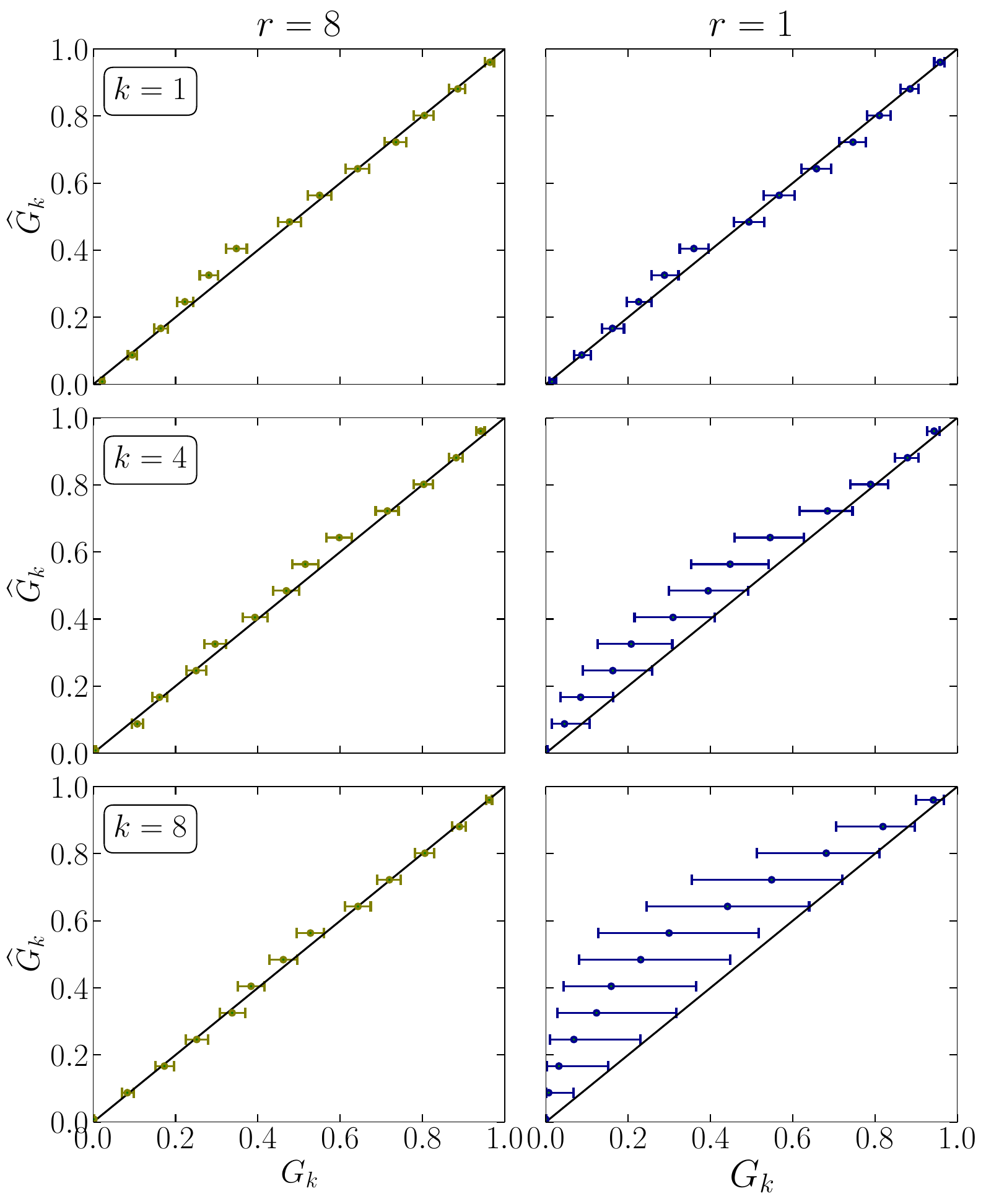}
    \caption{Model diagnostics probability plots for the $k$-th largest order statistics with $k\in\{1,4,8\}$ and $V=(400\,h^{-1}\,{\rm Mpc})^3$. The $x$-axis represents the $k$-GEV, whereas the $y$-axis is the empirical CDF. The error bars correspond to the $1\sigma$ errors on the posterior distribution of $\Theta_{\rm GEV}$ that are propagated to the values of $G_k$. The diagonal lines represent the idealistic scenario in which the $k$-GEV is exactly equal to the empirical CDF. For clarity, only a representative subset of approximately twelve data points is shown.}
    \label{fig:Diagnostic}
\end{figure}

\begin{figure}[ht]
    \centering
    \includegraphics[width=1.0\linewidth]{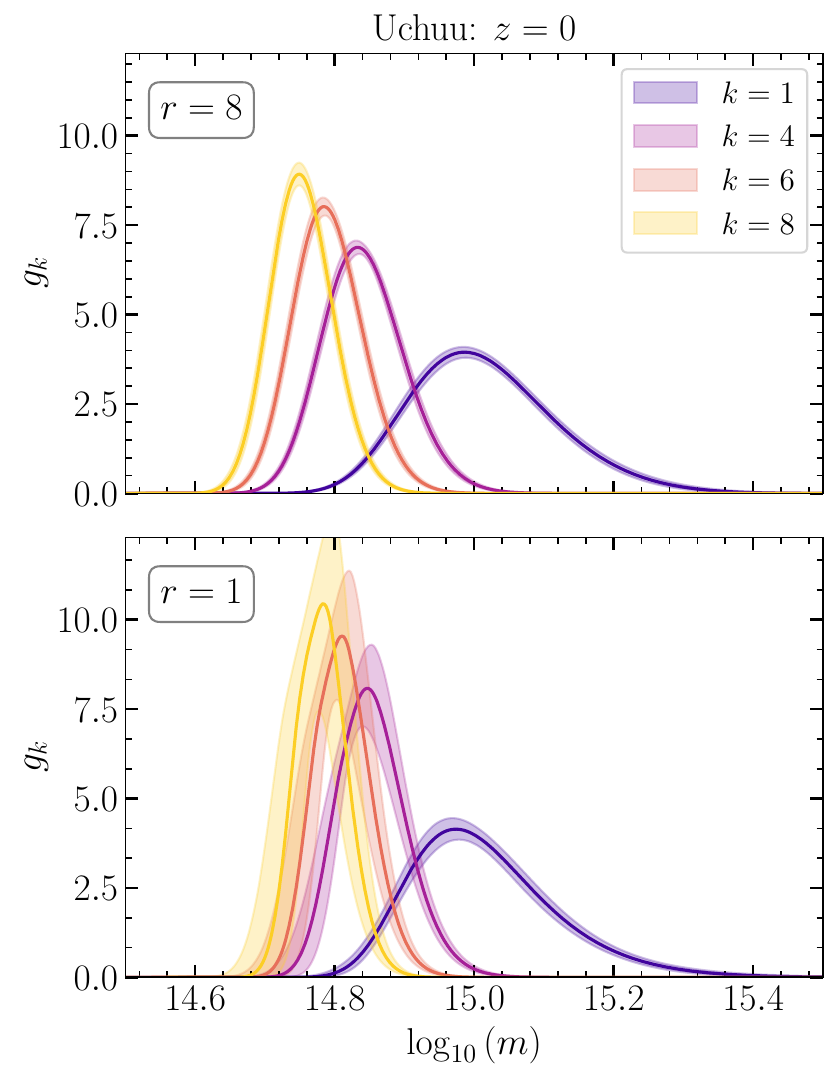}
    \caption{Comparison between the $k$-PDF estimated with the $r$-LOS approach for $k\in\{1,4,6,8\}$ using $r=8$ (upper panel) and $r=1$ (lower panel). The shaded regions denote the $68\%$ credible intervals derived from posterior samples, and solid lines correspond to the median values.}
    \label{fig:kpdf_comp}
\end{figure}

In Fig.~\ref{fig:Diagnostic}, we show probability plots of $ G_k(m_i^{(k)}) $ versus $ \hat{G}_k(m_i^{(k)}) $ for $ k = 1 $ (top panels), $k = 4$ (middle panels), and $k = 8$ (bottom panels), for the case $r = 1$ (right panels) and $ r = 8 $ (left panels). The data points represent the mean and $68\%$ credible intervals obtained by propagating the posterior uncertainty on the GEV parameters into Eq.~(\ref{kgev}). For clarity, only a subset of points is shown in Fig.~\ref{fig:Diagnostic}, as displaying all $125$ blocks would render the figure unreadable. For $k = 1$, both methods accurately recover the empirical distribution, with the estimates lying along the diagonal. However, for $k > 1$, the standard EVS approach ($r = 1$) exhibits systematic deviations from the empirical distribution, whereas the $r$-LOS method ($r = 8$) remains consistent.
Moreover, the $r$-LOS approach yields smaller uncertainties on $G_k$, reflecting the additional information gained from ranked higher-order statistics. This trend is further illustrated in Fig.~\ref{fig:kpdf_comp}, which shows the top-$k$ GEV distributions for $k = 1$ (purple), $4$ (pink), $6$ (orange), and $8$ (yellow), computed using the best-fit GEV parameters obtained with  $r = 1$ (bottom panel) and $r = 8$ (top panel). In the former case, the distributions for $k > 1$ are dominated by large uncertainties and are poorly distinguishable. In contrast, the $r = 8$ results exhibit well-separated distributions with reduced uncertainties, demonstrating that incorporating higher-order ranked statistics improves the goodness of fit and tightens the constraints on the GEV parameters.

\subsubsection{HMF reconstruction validation}
Here, we first use the $r$-LOS approach to infer the GEV parameters from the rank-ordered halo masses measured across multiple sub-volumes of the Uchuu simulation, following the procedure described in Section~\ref{GEV_inf}. We then use these inferred parameters to test the reconstruction of the HMF through the GEV--HMF parameter relation, Eq.~(\ref{equivalence_HMF}).

\begin{figure*}
    \centering
    \includegraphics[width=1.0\linewidth]{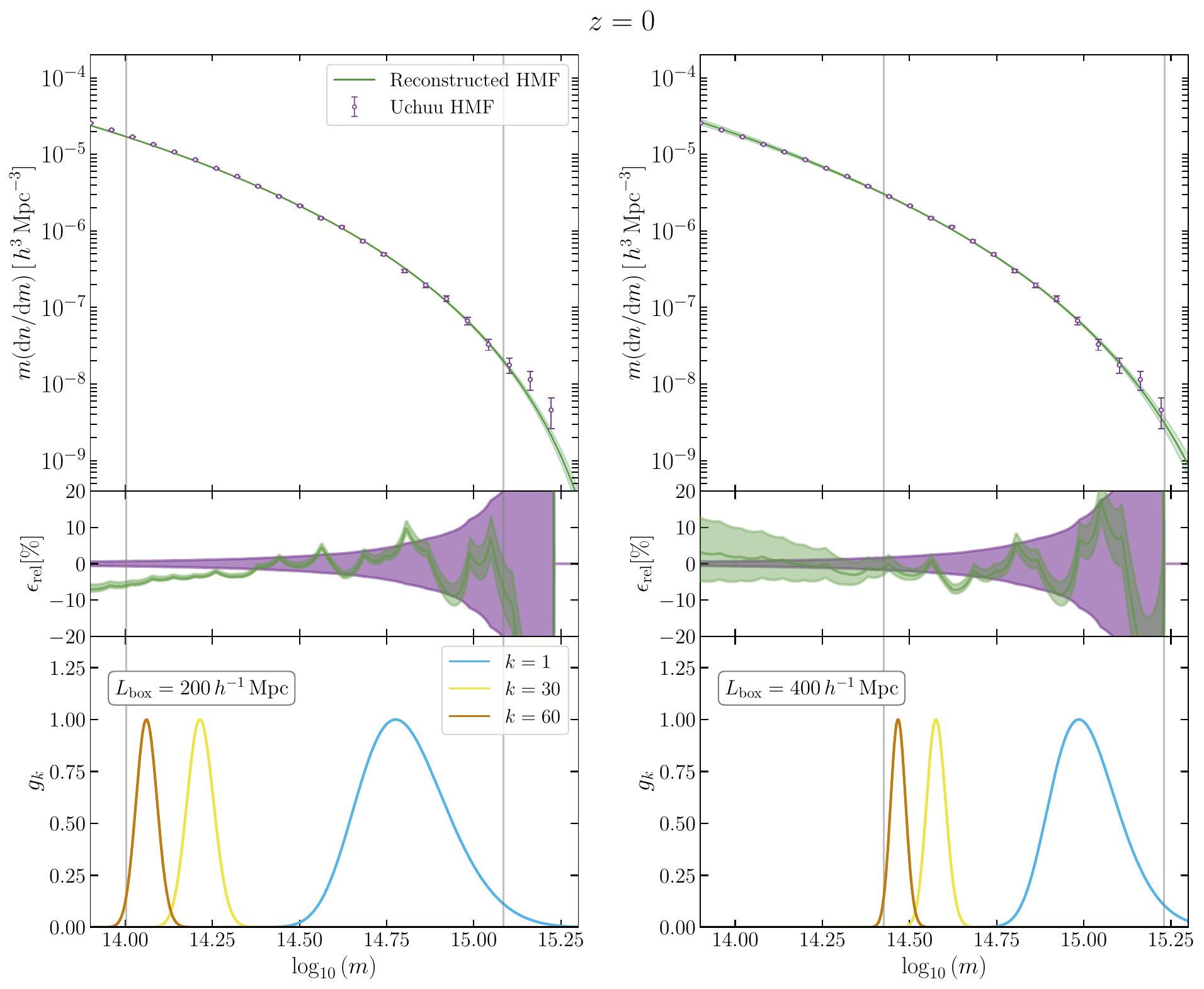}
    \caption{Reconstruction of the Uchuu-HMF at large halo masses for two different block sizes, $L_{\rm box} = 200\,{\rm Mpc}\,h^{-1}$ (left) and  $L_{\rm box} = 400\,{\rm Mpc}\,h^{-1}$ (right). The lower panels represent the distribution of the $k$-th order for $k=1$ (blue), $k=30$ (yellow) and $k=60$ (orange) using $r=60$. The vertical lines show the $2.5$-th and $97.5$-th percentiles for both rank $k$.
    The upper panels show the numerical estimate of the Uchuu-HMF at $z=0$ (purple dots) with their Poisson error bars against the reconstructed HMF (solid green lines). }
    \label{fig:simu_reconstruction}
\end{figure*}

Specifically, we perform two separate analyses: one using $N_V = 1000$ blocks with box size $L_{\rm box} = 200\,{\rm Mpc}\,h^{-1}$, and another using $125$ sub-volumes with $L_{\rm box} = 400\,{\rm Mpc}\,h^{-1}$. In both cases, the $N_V$ blocks cover the parent volume and are used to sample the multivariate random vector of ranked masses up to $r = 60$. We then perform an MCMC analysis to sample the posterior distribution of the GEV parameters and propagate their uncertainties to the HMF through Eq.~(\ref{equivalence_HMF}). The resulting median prediction, together with the $16$th and $84$th percentiles defining the central $68\%$ credible interval, is shown in green in the upper panels of Fig.~\ref{fig:simu_reconstruction} for $z=0$.

The reconstructed HMF is compared to the exact HMF from the halo catalogue of the full Uchuu simulation volume, estimated using Eq.~(\ref{HMF_estimate_Uchuu}). The middle panels display the relative difference between the reconstructed and the numerical exact HMF, while the lower panels show the top-$k$ GEV distributions obtained using the best-fit GEV parameters, $\Theta_{\rm GEV}$.
For $L_{\rm box} = 200\,h^{-1}\,{\rm Mpc}$, the reconstructed HMF agrees with the numerical estimate within Poisson uncertainties over the mass range: $10^{14.3}\lesssim M_{500c}[M_{\odot}\,h^{-1}] \lesssim 10^{15.2}$, while noticeable discrepancies appear at lower masses $ M_{500c}\lesssim 10^{14.3}\,M_{\odot}\,h^{-1}$. For the larger volume, $L_{\rm box} = 400\,h^{-1}\,{\rm Mpc}$, the agreement is significantly improved, extending beyond the mass range directly sampled by the fit and remaining within Poisson uncertainties over more than one decade in mass: $10^{14.15}\lesssim M_{500c}[M_{\odot}\,h^{-1}] \lesssim 10^{15.3}$.
These results indicate that: 1) for the larger block size, it is possible to choose an arbitrary high value of $r$ (i.e. $r\sim60$ for $L_{\rm box} = 400\,h^{-1}\,{\rm Mpc}$) without biasing the reconstruction due to the halo clustering within and between boxes; 2) it is possible to extrapolate the reconstruction beyond the empirical support of the fit for a well-chosen value of $r$ as shown in the right panel.

Overall, this suggests that selecting subsamples of the most massive clusters across multiple sub-volumes, where individual mass measurements are feasible, provides an efficient and robust method to reconstruct the high-mass end of the HMF. For example, in the full volume there are $N_h = 40\,800$ detected halos in the mass range $10^{14.15}\lesssim M_{500c}[M_{\odot}\,h^{-1}] \lesssim 10^{15.3}$, while we have only used $N_h = 7\,500$ halos to reconstruct the HMF in the same mass range, corresponding to $\sim 20\%$ of the total halo sample. 

\begin{figure*}
    \centering
    \includegraphics[width=1.0\linewidth]{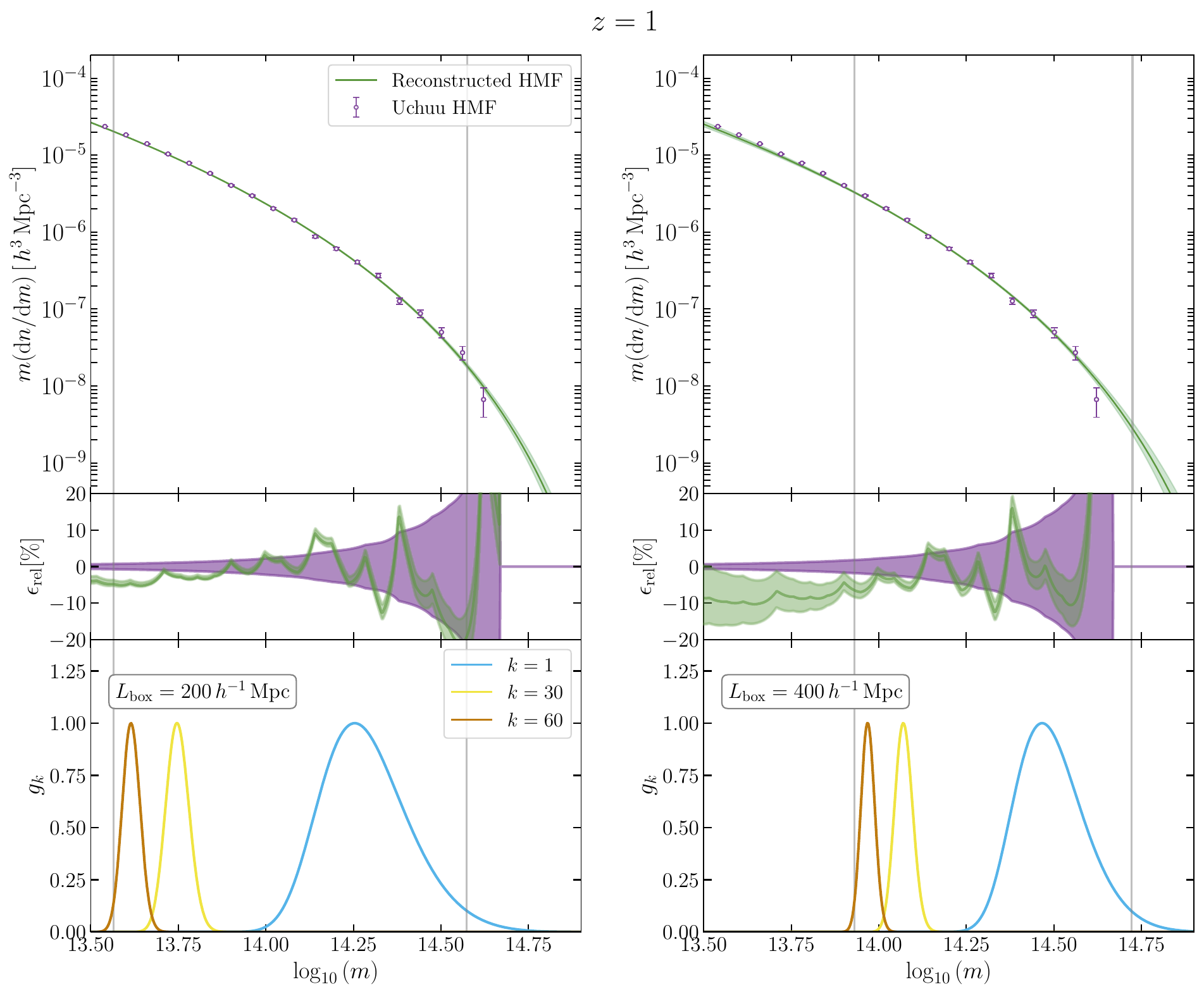}
    \caption{Same as Fig.~\ref{fig:simu_reconstruction} for $z=1$.}
    \label{fig:simu_reconstruction_z1}
\end{figure*}

We carried out the same analysis for the $z=1$ snapshot, the associated results are shown in Fig.~\ref{fig:simu_reconstruction_z1}. In the case of blocks with size $L_{\rm box} = 200\,h^{-1}\rm Mpc$, the reconstructed HMF agrees with the numerical estimate within Poisson uncertainties over the mass range: $10^{13.8}\lesssim M_{500c}[M_{\odot}\,h^{-1}] \lesssim 10^{14.6}$, while discrepancies appear for $ M_{500c}\lesssim 10^{13.8}\,M_{\odot}\,h^{-1}$. For the larger volume, $L_{\rm box} = 400\,h^{-1}\,{\rm Mpc}$, the reconstruction remains within the Poisson uncertainties in the range $10^{14.1}\lesssim M_{500c}[M_{\odot}\,h^{-1}] \lesssim 10^{14.75}$ while discrepancies appear for $M_{500c}\lesssim 10^{14.1}M_{\odot}\,h^{-1}$.

\subsection{Flagship redshift space analysis}
\subsubsection{Sampling the $r$-LOS multivariate random vector}
Similarly to the comoving space analysis discussed in Section~\ref{GEV_inf}, we aim to estimate the parameters of the GEV distribution of halos in redshift space using the $r$-LOS approach, which we use to reconstructed the number counts using Eq.~(\ref{number_counts}) and test it against the direct counting. To this end, we sample the multivariate random vector of the $r$ most massive halos within sub-volumes defined through the sky-tiling method introduced in \cite{waizmann_novel_2011}. These sub-volumes consist of spherical shells of fixed redshift width, $\Delta z$, spanning a solid angle $\delta\Omega$. 

To ensure the validity of the $r$-LOS formalism, the resulting light-cone patches must be sufficiently large, both in angular extent and depth. This guarantees that each patch contains a large number of halos, allowing for a reliable approximation of the asymptotic regime, while minimizing the impact of halo clustering both within individual patches and across different patches.

In practice, we use the \texttt{HEALPix}\footnote{\protect\url{http://healpix.sourceforge.net}} (Hierarchical Equal Area isoLatitude Pixelization) Python library \cite{Zonca2019,2005ApJ...622..759G} to partition the sky into $N_{\rm pix}$ equal-area pixels. First, the comoving Cartesian coordinates of the halos in the Flagship light-cone are converted into spherical coordinates, with the observer placed at the origin. We then set the \texttt{HEALPix} resolution parameter $N_{\rm side}$, which determines the subdivision of the sphere into $12 \times N_{\rm side}^2$ equal-area pixels. 

Next, we compute $N_{\rm octant}$, the number of pixels contained within the Flagship octant. To avoid biases from partially covered regions, we discard pixels whose area extends beyond the light-cone footprint. In practice, we retain only patches containing at least $60\%$ of the average number of clusters per patch.

For each patch, we construct the random vector $\mathbf{M}^{(r)}$, containing the $r$ most massive clusters, ranked in descending order in the $i$-th sub-volume. These data vectors are then used to estimate the GEV parameters via Eq.~(\ref{posterior_GEV}).

\subsubsection{Enhanced GEV parameter inference with $r$-LOS}
\begin{figure}[ht]
    \centering
    \includegraphics[width=1.0\linewidth]{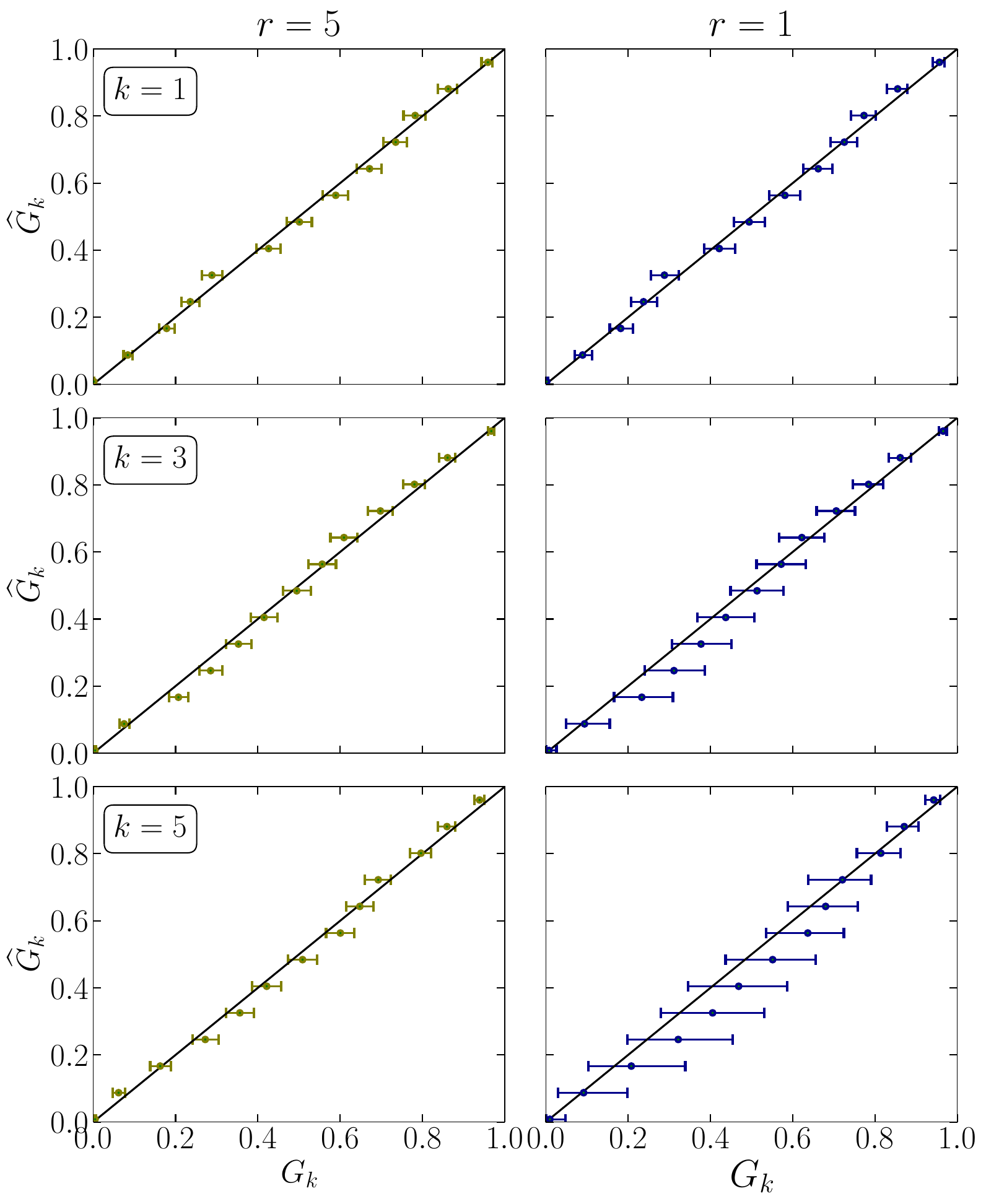}
    \caption{As in Fig.~\ref{fig:Diagnostic} for the halos in the Flagship light-cone catalog where $z\in[0.3,0.6]$ and $\delta\Omega \simeq 41\, {\rm deg}^2$.}
    \label{fig:Diagnostic_Flagship}
\end{figure}

As in the case of the diagnostic analysis of the Uchuu catalogs shown in Fig.~\ref{fig:Diagnostic}, in Fig.~\ref{fig:Diagnostic_Flagship} we show probability plots which confront the predicted values of the top-$k$ GEV distribution, $G_k$, assuming the GEV parameters inferred using the $r$-LOS approach for $r=5$ (left panels) and $r=1$ (right panels), against the empirical values for $k=1$ (top panels), $k=3$ (middle panels) and $k=5$ (bottom panels) for $z\in[0.3,0.6]$ and $\delta\Omega \simeq 40\, {\rm deg}^2$. We can see that the predicted values lie along the diagonal line for $r=5$ and for all values of $k$, while the case $r=1$ only works for $k=1$. For higher values of $k$, discrepancies and larger errors appear. 
In Fig.~\ref{fig:Diagnostic_Flagship}, we plot the corresponding PDFs of top-$k$ GEV distributions obtained using the GEV parameters for $r=1$ (bottom panel) and $r=5$ (top panel). As expected, we can see that the use of the higher-ranked masses of the $r$-LOS method significantly reduces the uncertainties on the estimated PDFs. Overall, this shows that the $r$-LOS methods provide a statistically robust estimate of the asymptotic form of the GEV distribution, both in comoving and redshift space.
\begin{figure}[ht]
    \centering
    \includegraphics[width=1.0\linewidth]{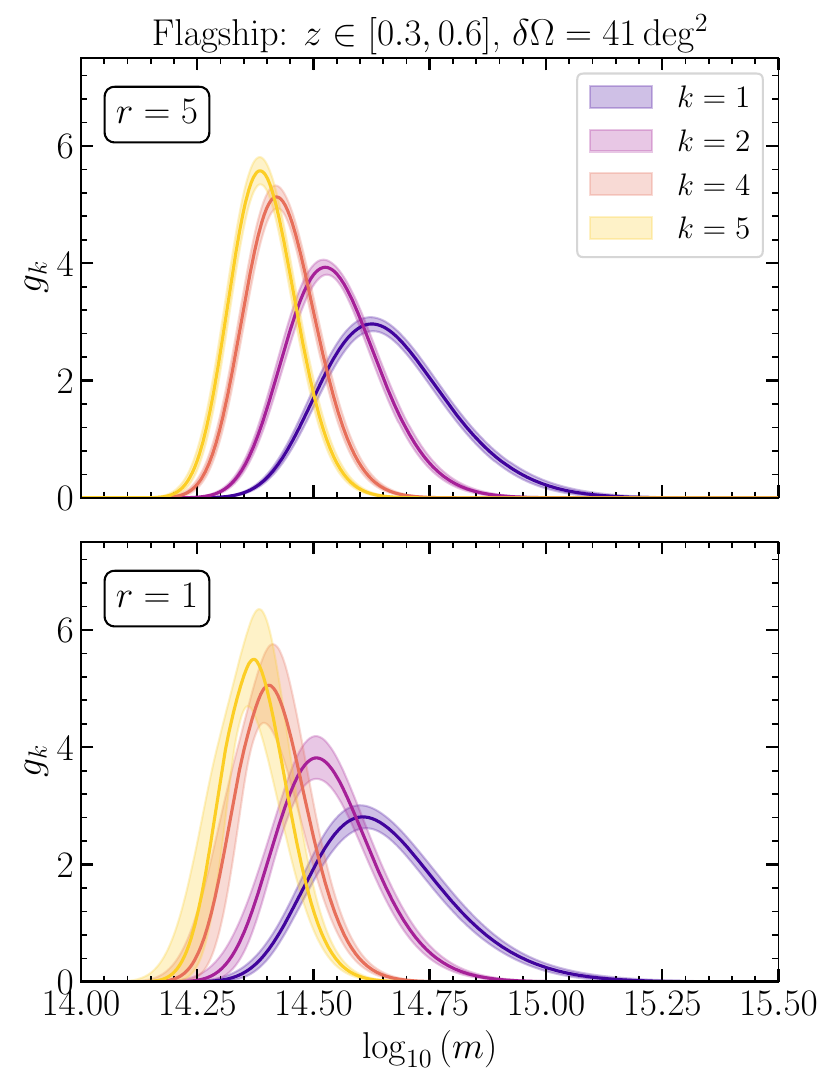}
    \caption{As in Fig.~\ref{fig:kpdf_comp} for the halos of the Flagship light-cone catalog in the redshift range $0.3\le z\le 0.6$. Comparison between the $k$-PDF estimated with the $r$-LOS approach for $k\in\{1,2,4,5\}$ using $r=5$ (upper panel) and $r=1$ (lower panel) respectively. The shaded regions denote the $68\%$ credible intervals derived from posterior samples, and solid lines correspond to the median values.}
    \label{fig:kpdf_comp_Flagship}
\end{figure} 

\subsubsection{Number counts reconstruction validation}
\begin{figure*}
    \centering
    \includegraphics[width=1.0\linewidth]{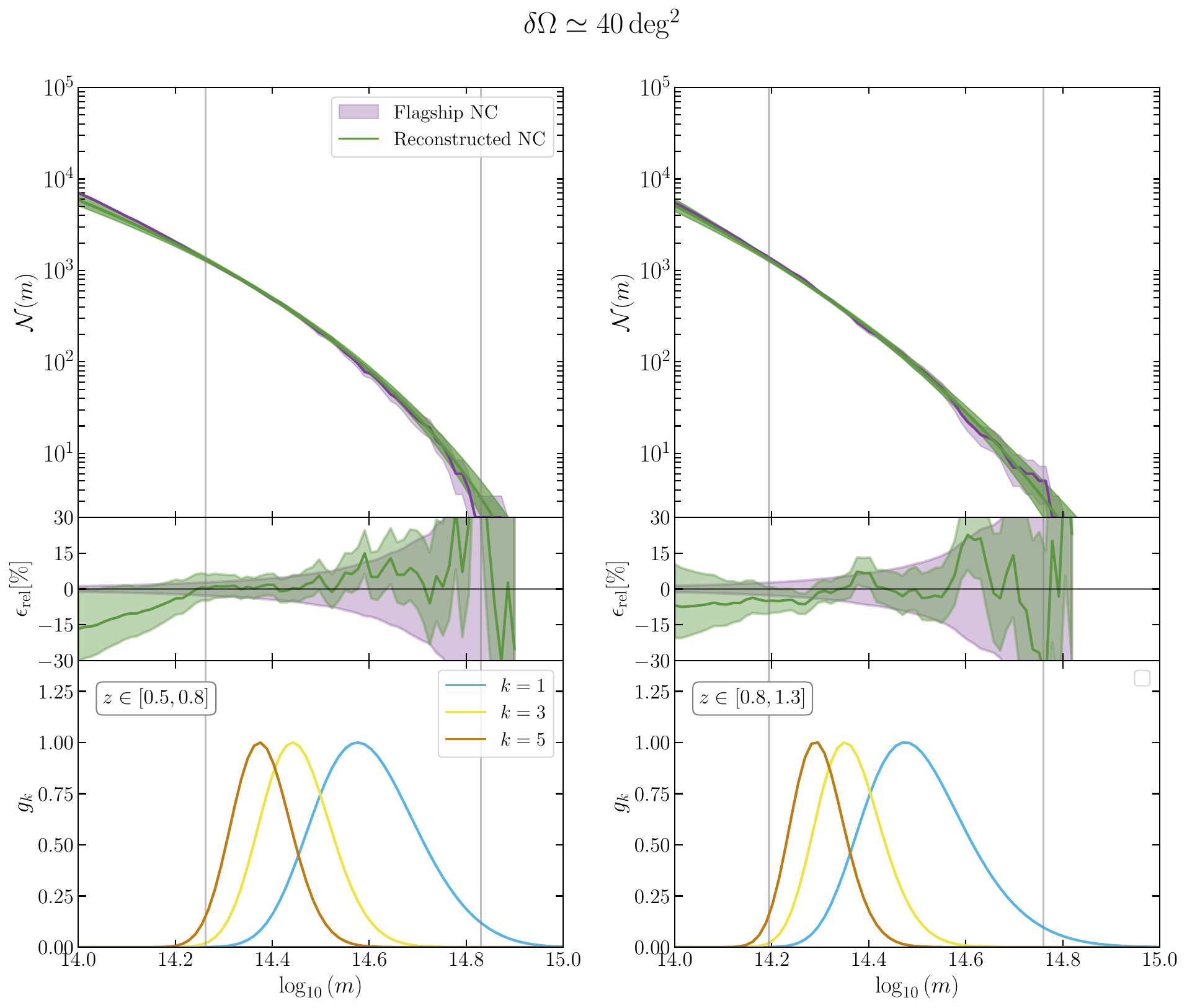}
    \caption{Reconstruction of the Flagship number counts for two different redshift slices $z\in[0.5,0.8]$ (left) and  $z\in[0.8,1.3]$ (right), for patches covering a sky area of $\delta\Omega_{\rm sky} \simeq 40\,\deg^2$. The lower panels represent the distribution of the $k$-th most massive clusters for $k=1$ (blue), $k=3$ (yellow), and $k=5$ (orange). The dashed lines correspond to the median values of the posterior samples of $\Theta_{\rm GEV}$. The vertical lines show the $2.5$-th percentile of the $5$-GEV distribution and the $97.5$-th percentile of the $1$-GEV distribution.
    The upper panels show the Flagship number counts (solid black line) with their Poisson error bars (gray shaded region). The middle panel shows the relative error between the reconstructed and the Flagship number counts. }
    \label{fig:simu_reconstruction_NC}
\end{figure*}

Similarly to the comoving space analysis, we reconstruct the Flagship number counts for two distinct survey configurations covering $40$ deg$^2$ of the sky, one extending over the redshift interval $0.5\le z \le 0.8$ and the other over the range $0.8\le z \le 1.3$. 
The results are shown for the two cases in the left and right panels of Fig.~\ref{fig:simu_reconstruction_NC}, respectively. In the upper panels we plot the Flagship number counts (purple solid line) with the associated Poisson uncertainties (purple shaded band) against the GEV reconstruction (green solid line) with the associated uncertainties (green shaded band). In the middle panels, we plot the relative differences, while the bottom panels show the top-$k$ GEV distributions for $k=1,3$ and $5$ computed assuming the best-fit $\Theta_{\rm GEV}$ values. The vertical lines correspond to the mass range of the $r$-LOS fits. We can see that in both cases and within the mass range delimited by the vertical lines, we can retrieve the Flagship number counts to great accuracy. 

\section{Primordial non-Gaussianity on $r$-LOS of clusters}\label{sec:PNG}
The abundance of the most massive galaxy clusters has long been recognized as a sensitive probe of primordial non-Gaussianity, as deviations from Gaussian initial conditions can significantly alter the statistics of rare, high-density peaks \cite{1988ApJ...330..535L,matarrese2000,verde2000,loverde2008,Achitouv2012}. In this context, EVS provides a natural framework to quantify the likelihood of observing the most massive systems in a given survey volume. Several studies have explored the use of EVS to constrain PNG by comparing the observed most massive clusters with theoretical predictions for different non-Gaussian initial conditions (e.g., \cite{dalal2008,harrison_exact_2011,Hotchkiss_2011}). These analyses have primarily focused on the maximum halo mass and its probability distribution, demonstrating that positive (negative) PNG enhances (suppresses) the abundance of extreme objects. Although current observational constraints remain broadly consistent with Gaussian initial conditions, these results highlight the sensitivity of extreme cluster statistics to PNG. They also motivate the development of more refined approaches that exploit the additional statistical information encoded in multiple extreme objects, rather than relying solely on the most massive one.

To this end, we investigate the impact of PNG on the top-$k$ distribution of massive clusters, focusing in particular on local-type primordial non-Gaussianity, which produces the strongest deviations in the high-mass tail of the HMF.

\subsection{Quijote PNG simulations}
We use $300$ independent realizations of the Quijote-PNG simulations \citep{Quijote-PNG,2022ApJ...940...71J}, each covering a volume of $(1\,\mathrm{Gpc}\,h^{-1})^3$ and containing $512^3$ particles. The simulations explore three values of the local PNG parameter that are $f_{\rm NL}^{\rm local}\in\{-100,0,100\}$. We note that the latest \textit{Planck} DR4 constraints from bispectrum analyses \citep{Jung_2025} exclude these values. Our analysis should therefore be regarded as a proof of concept, aimed at demonstrating the sensitivity of the proposed method to PNG. The simulations adopt a fiducial cosmology with parameters $\Omega_{\rm m} = 0.3175$, $\Omega_{\rm b} = 0.049$, $h = 0.6711$, $n_s = 0.9624$, $\sigma_8 = 0.834$, $w = -1$, and $M_\nu = 0,\mathrm{eV}$. Halos are identified using the \texttt{Rockstar} halo finder \citep{2013ApJ...762..109B} and we analyze halo catalogues at $z=0.5$.

These simulations provide an ideal framework to investigate the impact of $f_{\rm NL}^{\rm local}$ on extreme objects, since each realization corresponds to an independent volume of box length $L_{\rm box} \gg 100\,\mathrm{Mpc}\,h^{-1}$, thereby naturally satisfying the conditions required for the application of the LOS formalism.

\subsection{PNG and GEV parameter inference}
We evaluate the PNG signature on the GEV parameters in the standard EVS approach with $r=1$ and in the $r$-LOS case with $r=15$. To this purpose, we sample the random vector of the $r$ most massive clusters in each realization of the Quijote simulations. From these samples, we infer constraints on the GEV parameters for the three values of $f_{\rm NL}^{\rm local}$ considered. 

\begin{figure}[ht]
    \centering
    \includegraphics[width=1.0\linewidth]{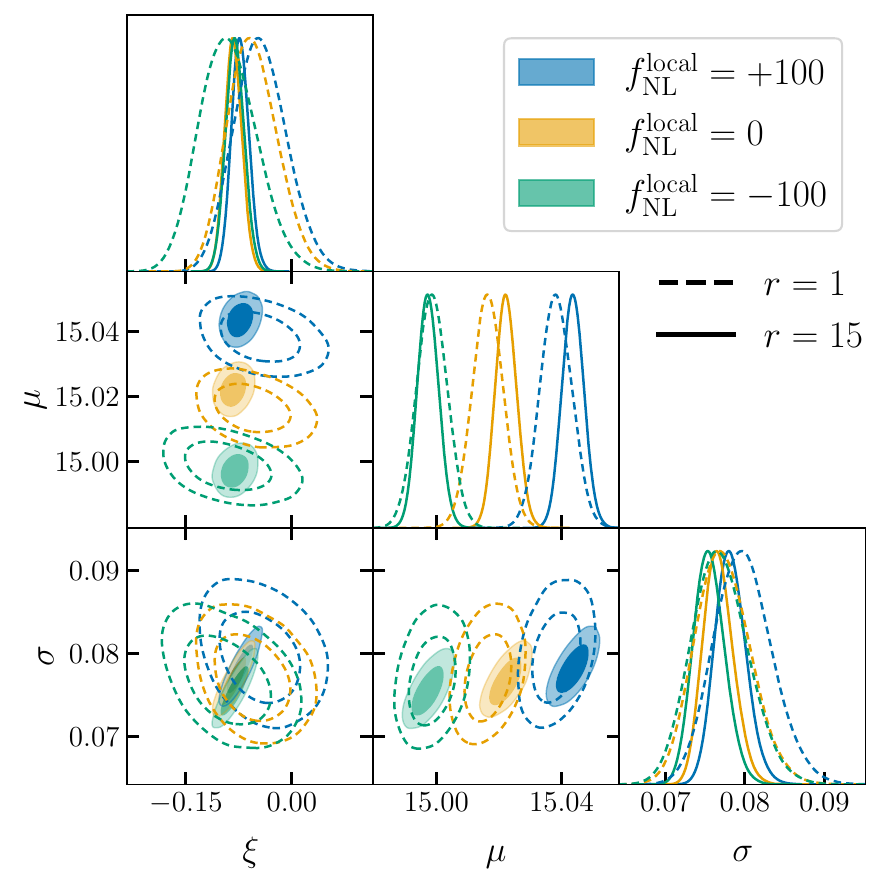}
    \caption{Triangular plot of the marginalized posterior distribution of the GEV parameters for the different $f^{\rm local}_{\rm NL}$ values at $z=0.5$, using $r=1$ (dashed empty contours) and $r=15$ (filled contours) halos per block}.
    \label{PNG_triangular_plot}
\end{figure}

Figure~\ref{PNG_triangular_plot} shows the posterior distributions of the GEV parameters inferred from the 300 simulations at $z=0.5$, using $r=1$ (dashed contours) and $r=15$ (filled contours). Including higher-ranked halo masses significantly tightens the posterior constraints compared to the standard block maxima approach ($r=1$), thereby enhancing the sensitivity of the inferred GEV parameters to variations in $f_{\rm NL}$. Focusing on the case $r=15$, we can see that the different PNG scenarios are most clearly distinguished by the location parameter $\mu$. The shape parameter $\xi$, by contrast, remains largely unaffected by negative values of $f_{\rm NL}^{\rm local}$, while positive values produce a slight shift toward higher values. This shift reflects the change in the slope of the HMF induced by the enhanced abundance of the most massive objects. The scale parameter $\sigma$ exhibits a different response: negative $f_{\rm NL}^{\rm local}$ values lead to a slight reduction in the spread of extreme masses, whereas positive values increase their scatter. 

These results highlight the potential of the \(r\)-LOS approach to significantly improve the sensitivity of extreme-value statistics to primordial non-Gaussianity by incorporating information from multiple ranked halos.

\section{Conclusions}\label{conclu}
The study of the most massive galaxy clusters provides a powerful and complementary avenue to test cosmology alongside more traditional galaxy cluster probes~\cite{allen2011,kravtsov2012}. In this context, the use of statistical tools specifically designed to characterize extreme events is essential for fully exploiting the cosmological information encoded in the high-mass tail of the halo population.

Extreme value statistics has been widely applied to galaxy cluster observations to assess the consistency of the standard cosmological model and to constrain cosmological parameters (e.g., \cite{Mortonson_2011,Harrison_2012,waizmann_modelling_2012,waizmann_order_2013}). In this work, we have developed a comprehensive extension of this framework based on the more general $r$-largest order statistics formalism. We derived the exact distributions of ranked halo masses in the asymptotic limit, demonstrating that the void probability function naturally emerges as a limiting case of the LOS approach. Building on the formalism introduced in \cite{coles_introduction_2001}, we established a direct connection between the GEV form of the LOS distributions and the HMF, leading to an estimator of the HMF valid for arbitrary rank $k$.

Within this framework, we presented a novel method to infer the GEV parameters from the $r$-LOS distributions. We showed that it provides a robust and efficient alternative to the standard block maxima approach. By incorporating information from multiple ranked objects, the $r$-LOS method significantly improves the precision of the inferred GEV parameters, in particular the shape parameter $\xi$, which is known to be extremely difficult to constrain, thereby enabling an accurate reconstruction of the top-$k$ distributions and, ultimately, of the high-mass end of the HMF. Our analysis shows that, within a finite volume, the high-mass tail of the HMF is well described by a GEV distribution with $\xi < 0$, implying the existence of an effective upper-mass limit, $\mathcal{M}^*$. This endpoint corresponds to a true upper mass limit only if the underlying HMF of the Universe belongs to the Weibull domain of attraction. We derive an exact expression for this effective mass endpoint, which enables its evaluation for arbitrary survey volumes and cosmological models.

Additionally, we investigated the impact of primordial non-Gaussianity on the statistics of extreme halos using dedicated non-Gaussian simulations. We showed that the inclusion of higher-order ranked statistics enhances the sensitivity to primordial non-Gaussianity compared to standard EVS analyses, opening new prospects for constraining deviations from Gaussian initial conditions.

Beyond the validation steps presented in this article, several directions could be explored to extend this work. We have shown that a subsample of massive clusters is sufficient to provide an accurate estimate of the high-mass tail of the HMF. Our method could therefore be used to calibrate HMFs from simulations that are computationally expensive and limited in spatial extent, such as those used to investigate alternative cosmological models involving inhomogeneous dark energy, modified gravity, or modified dark matter scenarios. In these cases, numerical estimates of the HMF obtained using the standard bin-counting methods are particularly limited at the high-mass end, where these bins are dominated by shot noise.
Furthermore, since a subsample of clusters is sufficient to constrain the high-mass end of the HMF, this opens up the possibility of directly inferring cosmological parameters from rank-ordered cluster samples. We leave this perspective for future work.

Overall, the $r$-LOS framework provides a promising statistical tool for exploiting the constraining power of forthcoming large-scale cluster surveys, such as \textit{Euclid}, \textit{eROSITA}, and future CMB experiments. By extracting information from multiple extreme objects rather than relying solely on the most massive cluster, it is particularly well-suited for the large statistical samples expected from next-generation surveys. An important next step will be to apply this methodology to observational data, accounting for mass uncertainties, selection effects, and survey systematics (e.g., \cite{mantz2010,pratt2019}). This will pave the way for fully exploiting extreme cluster statistics as a complementary and independent probe of cosmology.

\begin{acknowledgments}
TG thanks Stéphane Colombi, Siri Chongchitnan and Joe Silk for providing insightful comments on the manuscript. This work was supported by the Programme National Astro of
CNRS/INSU with INP and IN2P3, co-funded by CEA and CNES.
We thank Instituto de Astrofisica de Andalucia (IAA-CSIC), Centro de Supercomputacion de Galicia (CESGA), and the Spanish academic and research network (RedIRIS) in Spain for hosting Uchuu DR1, DR2, and DR3 in the Skies and Universes site for cosmological simulations. The Uchuu simulations were carried out on Aterui II supercomputer at Center for Computational Astrophysics, CfCA, of National Astronomical Observatory of Japan, and the K computer at the RIKEN Advanced Institute for Computational Science. The Uchuu Data Releases efforts have made use of the ${\rm skun@IAA\_RedIRIS}$, and ${\rm skun6@IAA}$ computer facilities managed by the IAA-CSIC in Spain (MICINN EU-Feder grant EQC2018-004366-P). This work has made use of CosmoHub, developed by PIC (maintained by IFAE and CIEMAT) in collaboration with ICE-CSIC. CosmoHub received funding from the Spanish government (MCIN/AEI/10.13039/501100011033), the EU NextGeneration/PRTR (PRTR-C17.I1), and the Generalitat de Catalunya. Some of the results in this paper have been derived using the \texttt{healpy} and \texttt{HEALPix} packages. For visualization purposes, we used the \texttt{Py-ART} python package~\citep{helmus2016pyart}. Anthropic's Claude and ChatGPT were used to assist with proofreading, debugging code, and suggesting code optimizations and improvements to plots.
\end{acknowledgments}
\appendix

\section{Asymptotic limit of the exact top-$k$ GEV distribution}\label{appendix_asym_k_gev}
We derive the large $N$-limit of the exact top-$k$ GEV distribution along the line of the work presented in \cite{borgman_frequency_1961}.

Let $\mathcal{N}(m)= n(>m) V$ be the number of halos with mass $>m$ in the volume $V$ and $N=n_{tot}V$ the total number of halos in the same volume. Then, we can rewrite Eq.~(\ref{fm}) as: 
\begin{equation}\label{fmbis}
F(m)=1-\frac{\mathcal{N}(m)}{N},
\end{equation}
and substituting into Eq.~(\ref{exact_rank}), we have
\begin{equation}\label{pithn2}
\Phi_{N}^{(k)}(m)=\left(1-\frac{\mathcal{N}(m)}{N}\right)^N\sum_{l=0}^{k-1}\binom{N}{l}\left(\frac{\frac{\mathcal{N}(m)}{N}}{1-\frac{\mathcal{N}(m)}{N}}\right)^l.
\end{equation}
In the large $N$-limit, the first factor in Eq.~(\ref{pithn2}) has for limit:
\begin{equation}
    \lim_{N\to+\infty}\left(1-\frac{\mathcal{N}(m)}{N}\right)^N = e^{-\mathcal{N}(m)},
\end{equation}
while the argument of the sum in Eq.~(\ref{pithn2}) reads as:
\begin{equation}
\binom{N}{l}\left(\frac{\frac{\mathcal{N}(m)}{N}}{1-\frac{\mathcal{N}(m)}{N}}\right)^l =
\frac{\left(1-\frac{1}{N}\right)..\left(1-\frac{l-1}{N}\right)}{l!\left(1-\frac{\mathcal{N}(m)}{N}\right)^l}\mathcal{N}(m)^l.
\end{equation}
Hence in the large $N$-limit this tends to $\mathcal{N}(m)^l/l!$. Therefore, the asymptotic form of the exact top-$k$ GEV distribution reads as:
\begin{equation}
\Phi^{(k)}(m)=e^{-\mathcal{N}(m)}\sum_{l=0}^{k-1} \frac{\mathcal{N}(m)^l}{l!}\equiv \mathcal{Q}(k,\mathcal{N}(m)).
\end{equation}

\section{GEV parameters and HMF relation}\label{appendix_peak}
We present the derivation of Eq.~(\ref{HMF_to_xi})-(\ref{HMF_to_mu}) following the approach adopted in \cite{davis_most_2011,waizmann_novel_2011}.

Let us perform a Taylor expansion of the GEV distribution $G$, as given by Eq.~(\ref{gev_blockmaxima}), and of the void probability distribution $P_0$, as given by Eq.~(\ref{void_distrib}). About the peak $m_0$, we have: 
\begin{align*}
    G(m) &= G(m_0)+\frac{1}{1!}\left.\frac{dG}{dm}\right|_{m_0}(m-m_0) \\
    &\quad + \frac{1}{2!}\left.\frac{d^2G}{dm^2}\right|_{m_0}(m-m_0)^2 +\, ...\\
    P_0(m)  &= P_0(m_0)+\frac{1}{1!}\left.\frac{dP_0}{dm}\right|_{m_0}(m-m_0) \\
    &\quad + \frac{1}{2!}\left.\frac{d^2P_0}{dm^2}\right|_{m_0}(m-m_0)^2 +\, ... 
\end{align*}
Evaluating these expressions at the peak value, we have $G(m_0)=P_0(m_0)$, that implies  $\tau(m_0)=\mathcal{N}(m_0)$, from which we obtain:
\begin{equation}
    \left[1 + \xi\left(\frac{m_0-\mu}{\sigma}\right) \right]^{-1/\xi} = \mathcal{N}_0. \label{first_eq}
\end{equation}
where $\mathcal{N}_0 \equiv \mathcal{N}(m_0)$ denotes the number of clusters with mass $m>m_0$. Equating the first-order terms of the $G(m)$ and $P_0(m)$ expansions at $m_0$ that is equivalent to evaluating Eq.~(\ref{equivalence_HMF}) at $m_0$, we further have:
\begin{equation}
\frac{1}{\sigma V}\left[1+\xi\left(\frac{m_0-\mu}{\sigma}\right) \right]^{-1/\xi -1} = \frac{dn(m_0)}{dm}, \label{second_eq}
\end{equation}
where we have used the explicit form of $\tau(m_0)$ in the case of $\xi\ne 0$. Notice that $m_0$ can be written in terms of the GEV parameters:
\begin{equation}\label{m_0}
    m_0 = \mu + \frac{\sigma}{\xi}\left[(1+\xi)^{-\xi}-1 \right].
\end{equation}
Substituting Eq.~(\ref{m_0}) into Eq.~(\ref{first_eq}) and Eq.~(\ref{second_eq}) and rearranging the GEV parameters on the left-hand-side we find 
\begin{align}
     \xi &= \mathcal{N}_0-1,\\
     \sigma &= \frac{-(1+\xi)^{1+\xi}}{\mathcal{N}'({m_0})},\\
     \mu &= m_0- \frac{\sigma}{\xi}\left[(1+\xi)^{-\xi}-1 \right]. 
\end{align}
The value of $m_0$ can be deduced by equating the second-order derivatives of $G(m)$ and $P_0(m)$ at the peak location, which yields the following equation:
\begin{equation}\label{root}
    I'(m_0)+V\,I^2(m_0) = 0,
\end{equation}
where $I(m_0)\equiv {\rm d}n/{\rm d}m|_{m_0}$ and $I'(m_0)={\rm d}I/{\rm d}m|_{m_0}$. This can be generalized to a survey volume covering a sky area $\Delta\Omega$ and a redshift interval $z\in[z_{\rm min},z_{\rm max}]$. In such a case, we find:
\begin{equation}
    \mathcal{I}'(m_0)+\Delta\Omega\,\mathcal{I}^2(m_0)=0,
\end{equation}
where 
\begin{equation}
    \mathcal{I}(m_0) =\int_{z_{\rm min}}^{z_{\rm max}}\frac{{\rm d}n}{{\rm d}m}\left(m_0,z\right)\frac{{\rm d}^2V}{{\rm d}z{\rm d}\Omega}\left(z\right)dz, 
\end{equation}
and $\mathcal{I}'(m_0)={\rm d}\mathcal{I}/{\rm d}m\vert_{m_0}$.

\bibliography{sample.bib}
\end{document}